\documentclass{emulateapj}
\usepackage{natbib} 
\usepackage{graphicx}
\usepackage{bm}
\usepackage{hyperref}
\usepackage{subfigure}
\usepackage{amsmath}
\usepackage{amssymb}
\usepackage{wasysym}
\usepackage{verbatim}

\def\rmw{\mathrm{w}}
\def\rmuw{\mathrm{uw}}
\def\ell{l}
\def\FeH{[\mathrm{Fe}/\mathrm{H}]}

\begin{document}

\title[Classically and Asteroseismically constrained 1D Stellar Evolution Models of {\Large $\alpha$} Centauri A and B using Empirical Mixing Length Calibrations]
{Classically and Asteroseismically constrained 1D Stellar Evolution\\ Models of {\Large $\alpha$} Centauri A and B using Empirical Mixing Length Calibrations}

\author{M. Joyce\altaffilmark{1}\altaffilmark{2} and B. Chaboyer\altaffilmark{1}}
\affil{$^{1}$Department of Physics and Astronomy,  Dartmouth College,  Hanover,  NH 03755}
\affil{$^{2}$South African Astronomical Observatory, PO Box 9, Observatory, Cape Town 7935, South Africa}

\begin{abstract}
The bright, nearby binary $\alpha$ Centauri provides an excellent laboratory for testing stellar evolution models, as it is one of the few stellar systems for which we have high-precision classical (mass, radius, luminosity) and asteroseismic ($p$-mode) observations. Stellar models are created and fit to the classical and seismic observations of both stars by allowing for the free variation of convective mixing length parameter $\alpha_{\text{MLT}}$.  This system is modeled using five different sets of assumptions about the physics governing the stellar models.  There are 31 pairs of tracks (out of ${\sim} 150,000$ generated) which fit the classical, binary, and seismic observational constraints of the system within  $3\,\sigma$. Models with each tested choice of input physics are found to be viable, but the optimal mixing lengths for {$\alpha$ Cen A and $\alpha$ Cen B} remain the same regardless of the physical prescription. The optimal mixing lengths are $\alpha_{\text{MLT,A}} /\alpha_{\odot} = 0.932$ and $\alpha_{\text{MLT,B}}/\alpha_{\odot} = 1.095$. That {$\alpha$ Cen A and $\alpha$ Cen B} require sub- and super-solar mixing lengths, respectively, to fit the observations is a trend consistent with recent findings, such as in \citet{Kervella17}, \citet{Joyce2018}, and \citet{Viani}. The optimal models find an age for $\alpha$ Centauri of $5.3 \pm 0.3$ Gyr.

\end{abstract}

\section{Introduction}
The $\alpha$ Centauri binary has long been a target of interest for the stellar modeling community because its proximity to Earth allows for high-precision observational constraints. Because both $\alpha$ Cen A and $\alpha$ Cen B are similar, but not identical, to the Sun in terms of fundamental parameters, the system also provides an excellent laboratory for testing the physical formalisms implemented in stellar models---many of which are extended from models of the Sun. Classical observations of $\alpha$ Cen span many decades, but recent interferometric observations by \citet{Kervella17} have given us new precision on the surface observables of both stars, including mass, radius, and luminosity.  Although there were previous attempts to obtain asteroseismic data for the $\alpha$ Cen system, detections of non-radial oscillations ($p$ modes) in these stars date back only to the work of \citet{BouchyCarrier} and \citet{Bedding2004}. Since these observations, asteroseismic constraints have been used to provide independent checks on many fundamental properties of the system, including radius, mass, age, and---most critically---the existence and nature of convective regions inside the stellar interior (see, for instance, \citealt{Bazot2016}, \citealt{Aguirre13}).  

Thanks to the observational work of \citet{deMeulenaer}, \citet{Kjeldsen}, and \citet{Bazot}, we now have a fair number of $p$-mode oscillation measurements in both {$\alpha$ Cen A and B.} From these frequencies, we can obtain the seismic parameter $r_{02}$, which has been shown by e.g.\  \citet{RoxVoronstov}, \citet{Aguirre13} to be a reliable probe of the stellar interior. This combination of rigorous classical and seismic constraints for both stars provides an ideal landscape for testing the validity and potential of stellar evolution models.

With increased access to high-precision empirical constraints, better use of the 1D convective mixing length parameter $\alpha_{\text{MLT}}$ \citep{EBV} has evolved as a focus in stellar modeling. Recognition of the inadequacy of using the solar-calibrated mixing length in models of other stars dates back to the mid-1980s, when it was noted by  \citet{Lattanzio84} and \citet{Demarque86} that the radius of $\alpha$ Cen A could not be reproduced without adjustments to $\alpha_{\text{MLT},\odot}$. This has been consistently corroborated since, such as by \citet{Fernandes95}, \citet{GD2000}, and \citet{Miglio05}.

Studies of binary systems have suggested that mixing length should depend on stellar mass in particular, such as \citet{LudwigSalaris99}, \citet{Morel00}, \citet{Lebreton01}, \citet{Lastennet}, \citet{Yildiz}, and \citet{Kervella08}. In addition, three-dimensional radiative hydrodynamic simulations of convection predict that the mixing length should also depend on the luminosity, metallicity, and surface gravity (e.g.\ \citealt{Freytag99}, \citealt{Ludwig99}, \citealt{Trampedach07}, \citealt{TrampedachStein11}, \citealt{Trampedach14}, and  \citealt{MagicMLT}).

From the asteroseismic perspective, \citet{Metcalfe10}, \citet{Deheuvels11}, and \citet{Mathur12} have shown that oscillations derived from stellar spectra cannot be reproduced using the solar mixing length. In particular, \citet{Metcalfe12} demonstrated that asteroseismic models of  16 Cyg A and B required non-solar values of $\alpha_{\text{MLT}}$ in both stars. Following this, \citet{Bonaca} found a positive correlation between mixing length and metallicity, a finding later corroborated by \citet{Tayar} in the context of \textit{Kepler} red giants. Between these studies, however, \citet{Aguirre15} performed  parameter-fitting algorithms which included asteroseismic constraints on a number of {\it Kepler} targets and found that the impact of varying $\alpha_{\text{MLT}}$ was not significant, reporting that this effect was smaller than their statistical errors.

Since then, \citet{Creevey17}, \citet{Joyce2018},\citet{Viani}, and \citet{Liu2018} have presented results demonstrating the need for non-solar mixing lengths in robust stellar models, with \citet{Viani}, \citet{Creevey17}, \citet{Bonaca} providing equations which predict mixing length as a function of effective temperature, surface gravity, and metallicity.

In particular, \citet{Joyce2018} investigated the relationship between $\alpha_{\text{MLT}}$ and stellar evolutionary phase (main sequence, sub-giant, or red giant branch), using empirical data from five highly metal-poor stars and one globular cluster. By allowing the mixing length to vary in constructing best-fitting models to each of these targets, we found that sub-solar mixing lengths were necessary to achieve agreement with observations. \citet{Joyce2018} indicates that stars with well-constrained physical properties can be used to calibrate the mixing length used in stellar models. With its well-determined stellar properties and metal-rich composition \citep{PortoDeMello08}, $\alpha$ Cen provides an  ideal system for extending our previous work from very metal-poor stars to metal-rich stars.

Previous modeling of $\alpha$ Centauri has taken many forms. Roughly a decade ago, \citet{Miglio} performed a parameter optimization on $\alpha$ Cen A and B using both classical and seismic constraints based on asteroseismic data from the very first observations of solar-like oscillations in the system. They did not find that fundamental stellar parameters depended on $\alpha_{\text{MLT}}$, but did find that such properties were sensitive to the treatment of diffusion in their models. \citet{Yildiz} found that the optimal mixing length value for models of $\alpha$ Cen A dropped by ${\sim}25$\% when the asteroseismic constraints were prioritized over classical constraints, and that prioritizing seismic agreement led to considerably lower age estimates for the system---8.9 Gyr classically versus 5.7 Gyr seismically.

\citet{Aguirre13} recognized the sensitivity of models' agreement to parameters which characterize the interior, finding that solutions having similar global properties using ``slightly different criteria to define what the best-fit model is can lead to solutions with similar global properties but very different interior structures'' \citep{Aguirre13}.

Soon after, \citet{Bazot2016} performed an impressive statistical simulation designed to uncover the probability that $\alpha$ Cen A has a convective core, using astrometric, spectroscopic, interferometric, and asteroseismic observations. Though their analysis does include constraints on the system derived from its binary attributes (e.g.\ astrometric measurements), it does not include optimization for $\alpha$ Cen B.  Via this method, they report a 40\% chance that $\alpha$ Cen A exhibits core convection. Using similar fitting analysis, which carefully considers many parameters, but which also fits {$\alpha$ Cen A} alone, \citet{Nsamba} recently report a 70\% chance of core convection in {$\alpha$ Cen A.} 

In this investigation, we simultaneously fit the parameters of $\alpha$ Cen A and B. In our fits, we allow for the free variation of $\alpha_{\text{MLT}}$ across a multi-dimensional grid of stellar models generated with the one-dimensional Dartmouth Stellar Evolution Program (DSEP) code. The grids span a host of input parameters for both stars and five different prescriptions for the modeling physics.  The mixing lengths in models of $\alpha$ Cen A and B  vary independently, while the tracks are required to fit the observational constraints at a common age for both stars. We present agreement statistics which include consideration of each star's individual fit to classical parameters, binary fit criteria, and asteroseismic consistency, for five sets of assumptions about the modeling physics. We find 31 models which fit all criteria and discover that they converge tightly to particular optimal mixing length values for {$\alpha$ Cen A and $\alpha$ Cen B.} 
{The $\alpha_{\text{MLT}}$ values relative to the solar-normalized mixing lengths (i.e. $\alpha_{\text{MLT}} / \alpha_{\odot}$) remain constant regardless of choice of input physics.}

The observational data is summarized in \S \ref{obs}. The general fitting approach is discussed in  \S \ref{models},  and \S \ref{modproc} presents the  model grid, parameter space sampled, and analysis of the best-fitting mixing lengths found with this procedure. \S \ref{astero} presents an analysis of the asteroseismic parameters, and in \S \ref{math}, we discuss the statistical methods used to determine classical and asteroseismic goodness-of-fit scores. The best-fitting models and their preferred parameter spaces are presented in \S \ref{results}.  We conclude in \S \ref{conc} with a presentation of the best-fitting parameters for $\alpha$ Cen uncovered with this technique, including mixing lengths for {$\alpha$ Cen A and $\alpha$ Cen B} found by empirical calibration and the age of the system.

\section{Classical and Asteroseismic Observations}
\label{obs}

$\alpha$ Cen A and B have been the subjects of extensive observations spanning many decades.  Most recently, \citet{Kervella17} used the near-infrared VLTI/PIONIER interferometer to measure the angular diameter of both.  These data were combined with parallax measurements of the system obtained by \citet{Kervella16} to determine the linear radii of the two stars. \citet{Kervella16} also determined the mass of the two stars from an analysis of the binary orbit and their luminosities from bolometric flux and parallax.  Thanks to the proximity of $\alpha$ Cen A and B, their radii and masses are determined with an uncertainty of $0.5\%$, while the luminosities are known to within $0.7\%$.  

The chemical composition at the surface of $\alpha$ Cen A and B was determined by \citet{PortoDeMello08}, who used high-resolution, very high $S/N$ spectra of the two stars in a differential abundance analysis (with respect to the Sun) to determine that the two stars are metal-rich. \citet{PortoDeMello08} present nearly identical abundances: $\FeH = +0.24\pm 0.03$ for {$\alpha$ Cen A} and $\FeH = +0.25\pm 0.04$ for {$\alpha$ Cen B.} These are converted to a ratio of the mass fraction of the heavy elements with respect to hydrogen ($Z/X$) { using  $Z/X_{\odot} = 0.022\pm 0.03$} from \citet{Grevesse} (see discussion in \citealt{Thoul03}). All of these observations provide us with information on the global properties of the stars (masses, radii, luminosities, and abundances), and we refer to these observations collectively as ``classical observations.''  

In addition to numerous classical campaigns, $\alpha$ Cen A and B have also been the targets of several observational campaigns designed to determine the frequencies of the non-radial acoustic pressure waves, or $p$-modes, at the surface of the stars.  These asteroseismic measurements carry information about the stellar interior. Modes of different harmonic degrees $l$ penetrate to different depths within the star, and their observed frequencies depend on the sound speed ($c$) in that region. The sound speed is a function of temperature ($T$) and composition (parameterized by the mean molecular weight, $\mu$), via $c^2  \propto T/\mu$,  and so the observed frequencies containinformation on the internal temperature and composition of the star.

Since all modes travel through the outer layers of the star, the observed
frequencies are affected by the complicated physics of the non-adiabatic
zone at the surface of the star. To first order, this effect can be
suppressed by examining differences in frequencies using the large and
small frequency separations, defined as
\begin{align}
\Delta_\ell (n) &= \nu_{n,\ell} - \nu_{n-1,\ell},
\\
d_{\ell,\ell+2} (n) &= \nu_{n,\ell} - \nu_{n-1,\ell+2},
\end{align}
respectively.
The observed frequency is given by $\nu$, $n$~is the radial order of
the harmonic, and $\ell$ refers to the harmonic degree, which, along
with azimuthal order $m$, characterizes the behavior of the mode over
the surface of the star. Definitions of these parameters vary slightly
from author to author, particularly regarding the starting index. The
formulae above reflect the format given in
 \citet{RoxVoronstov}, hereafter RV2003, and we adopt these definitions
 throughout our analysis.

The large frequency separation $\Delta_\ell(n)$ (hereafter abbreviated
$\Delta \nu$), on its own is a measure of the separation between
consecutive $p$-mode overtones and scales as the inverse of sound travel
time across the stellar diameter. This can serve as an independent
constraint on the stellar radius. The small frequency separation
(hereafter $d \nu$)  is sensitive to the gradient of sound speed in the
core.
As a result, the small frequency separation provides information
on the chemical composition gradient in this region. This can serve
as an independent constraint on the evolutionary phase of the star
\citep[e.g.][]{Aguirre13, DiMauro}.

Although the small and large frequency separations are designed to
suppress the influence of the surface layers on the seismic observables,
RV2003 demonstrate that the surface layers of the star still impact the
frequency separations.
To mitigate this further, they introduce the ratio of the small to large
separations, $r_{02}$, defined as
\begin{align}
r_{02} (n) = \frac{d_{0,2} (n)} { \Delta_1 (n)}.
\end{align}

RV2003 demonstrate that this ratio is unaffected by the physics of the surface layers, but rather depends solely on the conditions in the inner layers of the star. RV2003 also introduce two other frequency ratios which are commonly used in seismic analysis, however, these frequency ratios are not independent, from each other or from $r_{02}$. We elect to use $r_{02}$ alone in our analysis, as it has the smallest observational uncertainties.  

During the main sequence lifetime of a star, the change over time in the mean molecular weight due to hydrogen burning dominates over the slight increase in temperature \citep[e.g.][]{Floranes}, and the sound speed decreases. Since the masses of $\alpha$ Cen A and B are well constrained by the classical observations, $r_{02}$ depends primarily on the age of the star.  

The most recent seismic observational analysis of $\alpha$ Cen A is given by \citet{deMeulenaer}, who detected 44 modes with $\ell$ values ranging from 0 to 3 and $n$ values ranging from 15 to 28. Seismic observations of $\alpha$ Cen B were most recently obtained by \citet{Kjeldsen}, who detected 37 modes with $\ell = 0$--$3$ and \mbox{$n = 17$--$32$}. These data were used to determine the small and large frequency separations for A and B, respectively.  Classical and seismic observations of $\alpha$ Cen A and B are summarized in Table \ref{syspar}. 

\begin{table*} 
\centering 
\caption{ Observational Parameters }
\begin{tabular}{c l  l l l  }  
\hline\hline
& Property				&  $\alpha$ Cen A	 & $\alpha$ Cen B       & Reference \\ \hline
&Mass  $M_{\odot}$	    & $1.1055 \pm 0.004$ &   $0.9373 \pm 0.003$ & \citet{Kervella17} \\
&Radius $R_{\odot}$	    &  $1.2234\pm0.0053$ &   $0.8632\pm0.004$   & \citet{Kervella17} \\
&Luminosity $L_{\odot}$ & $1.521\pm0.015 $	 &  $0.503\pm0.007 $    & \citet{Kervella17} \\
&$Z/X$			        & $0.039\pm 0.006 $	 &  $0.039\pm 0.006 $   & \citet{PortoDeMello08}; \citet{Thoul03} \\
&$\Delta_{1}$			&  $105.9 \pm 0.3$	 &   $160.1 \pm 0.1$	& \citet{deMeulenaer}; \citet{Kjeldsen} \\
&$d_{02}$			    &  $5.8 \pm 0.1$	 & $10.7 \pm 0.6$	    & \citet{deMeulenaer}; \citet{Kjeldsen} \\
&$r_{02}$		        &  $0.055 \pm 0.001$ &    $0.066 \pm 0.004$ & \citet{deMeulenaer}; \citet{Kjeldsen} \\
 \hline
\end{tabular}
\tablecomments{Observed classical and seismic parameters of $\alpha$ Cen A and B. }
\label{syspar}
\end{table*}

\section{Stellar Modeling Procedure}
\label{models}

The (Dartmouth Stellar Evolution Program) DSEP code \citep{Chab06, Dotter} is a one-dimensional stellar evolution code which has demonstrated particular robustness in modeling low-mass stars. A thorough discussion of DSEP's mechanics is provided most recently in \citet{Dotter}, but some adjustments have since been implemented. Upgrades include updates to nuclear reaction rates \citep{Adel, Marta} and the ability to produce output suitable for  stellar oscillation analysis. This has been possible through the introduction of routines to track and organize asteroseismic parameters and produce ``gong'' files readable by oscillation codes such as GYRE \citep{GYRE}.

We use DSEP to generate two independent grids of stellar tracks, one tailored to $\alpha$ Centauri A and one to~B, as the two stars are widely separated and do not interact. Classical best-fitting models are determined based on  (1)~their ability to reproduce the known, common surface abundance $Z/X_{\text{surf}}$, and (2) their agreement with the observed luminosity, radius, and mass. Since the $\alpha$ Cen system {is assumed} to have formed from the same protostellar cloud, the two stars should have the same age and initial chemical composition.  This constraint is enforced by computing a closeness score based on agreement between input helium ($Y_{\text{in}}$) and metal ($Z_{\text{in}}$) abundances and the age at which the observable constraints (mass, radius, luminosity, and surface abundance) are satisfied.

Unlike for the Sun, there is no {model-independent method} which can determine the age of the $\alpha$ Cen system; rather, the age is estimated by stellar model fitting. We thus allow for the free variation of age, requiring only that independent models of $\alpha$ Cen A and $\alpha$ Cen B satisfy their respective classical constraints simultaneously (within grid resolution), with the restriction that the stars are not in the pre-main sequence phase of evolution. 

After isolating pairs of simultaneous best-fitting models by measure of agreement with classical observations,  high-resolution stellar tracks tuned to these parameters are calculated.  These high-resolution stellar structure models use a higher number of grid points and a more sophisticated equation of state \citep{FreeEOS}, and they are formatted to allow accurate asteroseismic frequency calculations using the GYRE code \citep{GYRE}. Theoretical large and small frequency spacings for each high-resolution track are computed over the relevant frequency ranges and $n$ modes observed for $\alpha$ Cen~A and $\alpha$ Cen B \citep{deMeulenaer,Kjeldsen}. 

In principle, the observations over-determine the stellar models. There are 9 observables---mass, radius, luminosity, and ratio of small-to-large frequency separations for each of $\alpha$ Cen A and B, along with the observed chemical composition ([Fe/H]) of the system---and seven adjustable input parameters for the stellar models---input mixing lengths, helium abundances, and heavy element mass fractions for each star, and the age of the system (though we also allow for the mass to fluctuate slightly, it does not vary drastically enough to be treated as adjustable). As such, there is no guarantee that models which fit the observations of $\alpha$ Cen A and B at a common age will be found. However, models which match all of the observational constraints are, in fact, found (see \S \ref{results} for a detailed discussion) when non-solar values of the mixing length are used.

In our grid of models, the mixing length parameter is allowed to vary. In
order to compare optimal mixing length values between models with different
choices of input physics, solar models are calculated for each of our
physical configurations.  Solar mixing length calibrations are performed
by adjusting the mixing length, initial helium abundance ($Y_{\text{in}}$),
and initial heavy element abundance ($Z_{\text{in}}$) until  a solar-aged model
reproduces the observed solar radius, luminosity, and surface abundance
$Z/X$ to better than 0.1\% accuracy. 

There are uncertainties associated with a variety of physical phenomena
which occur in stars, and stellar evolution calculations require
knowledge of the physics of plasmas at high temperatures. For example,
heavier elements gravitationally settle and diffuse relative to their
lighter counterparts, and the calculation of the diffusion coefficients
is uncertain at the ${\sim} 30\%$ level (e.g.\ \citealt{Thoul94}). To 
help account for uncertainties in our knowledge of the fundamental physics
of stars, five different grids of stellar models are constructed. Each
prescription differs from the others in its assumptions about atmospheric
as well as interior processes. 

DSEP's atmospheric boundary conditions can be specified from a number of
possibilities, including both analytical approximations and grid-based
data. The ``default'' prescription in DSEP is the PHOENIX model
atmospheres \citep{Hauschildt}, as these are the surface boundary
conditions shown to reproduce observations most effectively (for
temperatures up to 10,000 K and $\log g=5.5$; see \citealt{Joyce2018}).
However, the PHOENIX tables used for surface boundary conditions (BCs)
do not contain details on the structure of the atmosphere, and so cannot
be used in seismic calculations. Seismic compatibility requires that
DSEP use an analytic model atmosphere for surface BCs, and so 
the \citet{Eddingtonttau} approximation to the gray model
atmosphere is used in four of our physical configurations. The impact of
choice in model atmosphere on our calculations is evaluated by computing
a grid of models which use \citet{KrishnaSwamy} surface
boundary conditions as well.  

As in \citet{Joyce2018}, we take into account variations in the efficiency of diffusion, denoted by the parameter $\eta_D$. We may consider $\eta_{\text{D}}$ to be a parameterization of the coefficients in the equations governing thermal diffusion and gravitational settling, as the treatment of diffusion in DSEP includes these two processes. Following the prescription of \citet{Thoul94}, H, He, and heavy elements are diffused, where heavy elements are represented as a single species assumed to diffuse at the same rate as fully ionized iron (see \citet{Chab2001} for a more involved discussion of diffusion formalisms in DSEP). It is sufficient to consider $\eta_D$ a measure of the diffusion rate, with higher values corresponding to shorter diffusion times. We consider cases with $\eta_D=1.0$, $\eta_D=0.5$, and $\eta_D=1.5$.  

The third major physical uncertainty in the models is the existence and nature of a convective core. While it is not thought that $\alpha$ Centauri B has a convective core, $\alpha$~Cen A's larger mass and the system's metal-enrichment  makes the existence of a convective core possible, and this has been the subject of many studies \citep[e.g.][]{Yildiz,Bazot2016, Nsamba}. While the majority of models of $\alpha$ Cen A generated with DSEP did not produce a convective core, core convection was occasionally activated when the models used enhanced diffusion ($\eta_D=1.5$).

The boundary between the convective core and radiative material above the core is determined using the standard Ledoux criterion, which assumes material stops instantaneously at the convective boundary.  This is not physically correct, and some amount of convective ``overshoot'' must occur.	In its default configuration, DSEP assumes that the convective overshoot is very small, and  the convective core overshoot parameter is set to zero.  To account for the uncertainties associated with determining the location of the convective core boundary, and the fact that it is permeable in physical reality, we generate a fifth model grid for $\alpha$ Cen A. This grid includes convective core overshoot in high-diffusion models, with an overshoot allowance of $\alpha_{\text{ovs}} = 0.1$, in units of pressure scale height ($H_p$).

Table \ref{solar} shows the calibrated mixing lengths and other key attributes for solar models generated under each of the five physical configurations considered. Solar models have a transient convective core near the zero-age main sequence, so convective core overshoot has a small effect on the solar-calibrated mixing length despite the lack of core convection in the Sun. The first column in Table~\ref{solar} provides shorthand signatures for the choices in input physics that will be referenced henceforth.

\begin{table*} 
\centering 
\caption{Theoretical Parameters of Solar-calibrated Models by Physical
Configuration}
\begin{tabular}{l lll  lll lll	}  
\hline\hline
 Config Name & Atmosphere & $\eta_{\text{D}}$ & $\alpha_{\text{ovs}}$ &
$\alpha_{\odot}$ & $Y_{\text{in}}$ & $Z_{\text{in}}$ &
 $\Delta\nu_{n,1}$  &$\delta \nu_{n,0}$ & $r_{02}$
\\ \hline
Standard				& Eddington    & 1.0 & 0.0 &
1.8210 &  0.27 & 0.018 & 135.4 & 9.85 &    0.0728 \\
KS						& Krishna Swamy & 1.0 &
0.0 & 2.1353 &	0.27 & 0.018 & 135.0 & 9.83 &	 0.0728 \\
Suppressed/Low Diffusion& Eddington    & 0.5 & 0.0 & 1.8148 &  0.28 &
0.020 & 135.6 & 9.89 &	  0.0729 \\
Enhanced/High Diffusion & Eddington    & 1.5 & 0.0 & 1.8535 &  0.27 &
0.018 & 134.6 & 9.67 &	  0.0718 \\
Overshoot				& Eddington    & 1.5 & 0.1 &
1.8559 &  0.27 & 0.018 & 135.2 & 9.68 &    0.0716 \\
 \hline
\end{tabular}
\tablecomments{ 
The large and small frequency spacings are averages taken over the range of harmonic $n$ values for which corresponding observations exist, in this case $n=9$--$26$ for solar seismic observations provided
 in \citet{Broomhall}. }
\label{solar}
\end{table*}

\section{Model Grids of $\alpha$ Centauri A and B }
\label{modproc}

Grids of stellar tracks for  $\alpha$ Centauri A and B  are calculated separately. The input parameter space includes variations in initial helium abundance $Y_{\text{in}}$, initial metal abundance $Z_{\text{in}}$, and mixing length $\alpha_{\text{MLT}}$. To account for uncertainty in mass, tracks for $\alpha$ Cen A can have 1.10, 1.105, or 1.11 $M_{\odot}$, and $\alpha$ Cen B tracks can have masses of 0.93 or 0.94 $M_{\odot}$. The other input parameters are sampled at the following initial resolutions:
\begin{itemize}
\item[] $\alpha_{\text{MLT}}: 1.0$ to $2.4$, $\delta_{\text{step}} = 0.1$.
\item[] $Y_{\text{in}}: 0.25$ to $0.45$, $\delta_{\text{step}} =0.01$.
\item[] $Z_{\text{in}}: 0.01$ to $0.046$, $\delta_{\text{step}} =${0.005}. 
\end{itemize}

\begin{figure}

\centering
\includegraphics[width=\columnwidth]{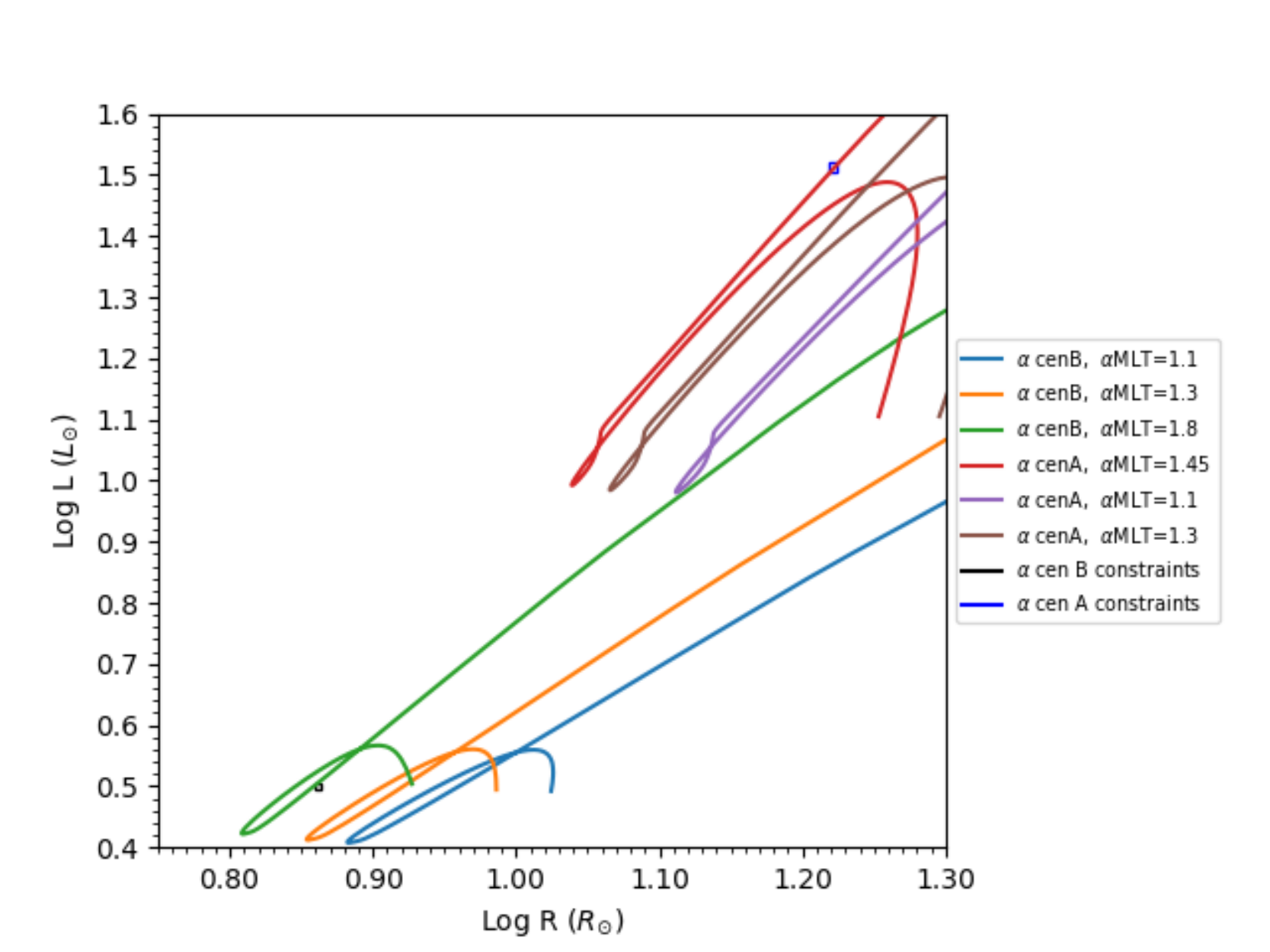}
\caption{ 
The three stellar tracks in the lower left quadrant are tested for $\alpha$ Cen B, one of which matches (green). In the upper right are three possible tracks for $\alpha$ Cen A, one of which matches (red). Boxed areas indicate observational constraints on radius and luminosity from \citet{Kervella17}. All models shown here are generated using initial abundances of $Z_{\text{in}}=0.025 $, $ Y_{\text{in}}=0.29$ for $\alpha$ Cen A and $Z_{\text{in}}=0.026 $, $ Y_{\text{in}}=0.28$ for $\alpha$ Cen B. Mixing lengths $\alpha_{\text{MLT}}$ for either star differ as indicated.
}
\label{acenAB}
\end{figure}

In our first sweep, we increase grid resolution iteratively on the parameter regions which are found to maximize the number of models that match the classical criteria.  A ``matching'' model is confirmed by (1) the intersection of the stellar track with the radius and luminosity given in \citet{Kervella17}, within a 2-D box defined by the observational uncertainty, and (2) agreement at the same point of intersection between the model star's surface abundance $Z/X$ and \citet{PortoDeMello08}'s abundance. Figure \ref{acenAB} shows a few stellar tracks which meet these criteria.

In cases where matching models emerge near the limits of the grid,we expand the grid in that direction. 

Adjustments to resolution and parameter limits are made independently for each star, and best-fitting parameters for one physical configuration do not influence the grid choices for any other prescription. The process continues until the mixing length reaches a resolution of $0.005$, $Y_{\text{in}}$ reaches a resolution of $0.01$, and $Z_{\text{in}}$ reaches a resolution of $0.001$.
Due to the case-specific refinement, the total number of stellar evolution tracks generated varies slightly per prescription. To achieve the necessary resolution, roughly 15,000 stellar tracks per configuration (times five prescriptions) are generated for $\alpha$ Cen~A, and roughly 18,500 stellar tracks per configuration (times four, as overshoot is not relevant in models without core convection) are generated for $\alpha$ Cen B. Our total grid consists of roughly $150,000$ stellar evolution tracks.

For each stellar track in the grid, the stellar structure equations are solved with tolerances of one part in $10^5$. An analytical equation of state which includes Debye--H\"uckel corrections is used. This yields fast execution times for the stellar evolution code, making it straightforward to run large grids of models. To limit computation time further, models are evolved only to the red giant branch, as both $\alpha$ Cen A and B lie on or just beyond the main sequence.  The grids are distributed over 16 cores and take roughly 20 hours to run at the initial parameter resolutions. Subgrids run on the order of a few hours. 

\subsection{Binary Constraints}

The definition of a model for $\alpha$ Cen as a system includes a pair of stellar tracks---a model each for {$\alpha$ Cen A and $\alpha$ Cen B}---and the definition of a ``fit'' or ``match'' becomes more rigorous as we eliminate inconsistent models and incorporate a greater number of constraints. For the most refined grids, we adjust parameter sampling according to an intermediate goodness-of-fit metric which formalizes agreement with the classical criteria sought initially: 
\begin{align}
s^2_{\text{classic}} & = \left [ \frac{R_{\text{A,obs}} -
R_{\text{A,mod}}}{\sigma_{R_{\text{A,obs}}} } \right ]^2  + 
 \left [ \frac{R_{\text{B,obs}} -
 R_{\text{B,mod}}}{\sigma_{R_{\text{B,obs}}} } \right ]^2 + \notag
 \\[3pt]
  & \;\;\; \;\; \left [ \frac{L_{\text{A,obs}} -
  L_{\text{A,mod}}}{\sigma_{L_{\text{A,obs}}} } \right ]^2  + 
 \left [ \frac{L_{\text{B,obs}} -
 L_{\text{B,mod}}}{\sigma_{L_{\text{B,obs}}} } \right ]^2 + \notag
 \\[3pt]
 &\;\;\; \;\; \left [ \frac{Z/X_{\text{obs}} -
 Z/X_{\text{mod}}}{\sigma_{Z/X_{\text{obs}}} } \right ]^2 ,
\label{classical_score}
\end{align}
where $R_{\text{A,obs}}$ is the observed radius of the {$\alpha$ Cen A} with uncertainty $\sigma_{R_{\text{A,obs}}}$,   $R_{\text{A,mod}}$ is the radius of the $\alpha$ Cen A model, $R_{\text{B,obs}}$ is the observed radius of $\alpha$ Cen B with uncertainty $\sigma_{R_{\text{B,obs}}}$, $R_{\text{B,mod}}$ is the radius of the $\alpha$ Cen B model, and similarly for the luminosity $L$ of both stars. The observed value of $Z/X$ for the $\alpha$ Cen system is $Z/X_{\text{obs}}$, with an associated uncertainty of $\sigma_{Z/X_{\text{obs}}}$, and $Z/X_{\text{mod}}$ is the average of the $Z/X$ values computed for the $\alpha$ Cen A and B models. 

While this statistic is useful for fine-tuning tracks to fit each star individually, determining the best-fitting pair of tracks according to this metric would neglect critical constraints imposed by binarity. Since $\alpha$ Centauri A and~B are members of the same system, we must take into account the fact that both theory \citep[e.g.][]{GoodKroup, Reipurth} and observation \citep[e.g.][]{King2012, Vogt2012, Mack2016, Liu2018} strongly suggest that two components of a binary system will have nearly the same age and {initial chemical composition.} 
We hence impose the requirement that viable models satisfy their respective observational constraints at a common age.  

\begin{figure}
\centering
\includegraphics[width=\columnwidth]{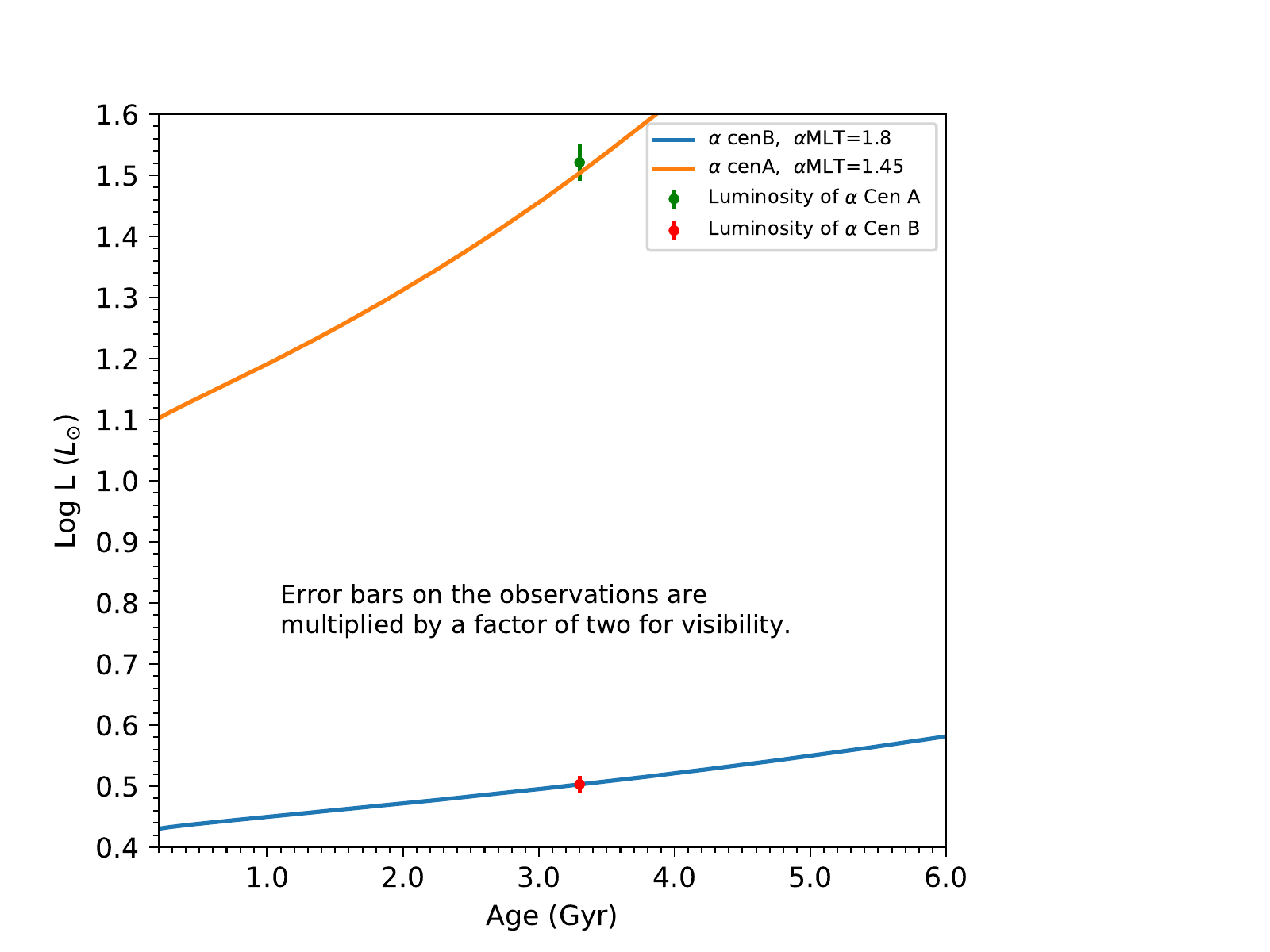} 
\caption{Stellar tracks for $\alpha$ Cen A (orange) and B (blue) are
shown intersecting their observed luminosities, indicated with green
and red markers, respectively, at a common age.}
\label{age}
\end{figure}

A classical match to the A and B system is thus any pair of tracks which each satisfy agreement with Table~\ref{syspar} at the same age, within a small, adjustable tolerance that is taken into account statistically (see \S \ref{math} for more detail). A model of $\alpha$ Cen as a whole comprises both {$\alpha$ Cen A and $\alpha$ Cen B models} (tracks). Figure \ref{age} demonstrates a candidate model for the system.

To account more properly for binary constraints, the final determination of best fit involves balancing the $s^2_{\text{classical}}$ score in equation \ref{classical_score} with a more robust binary agreement metric:
\begin{equation}
\medmuskip0mu
s^2_{\text{binary}}  =  \left [ \frac{\tau_A - \tau_B}{5\, \text{Myr}}
\right ]^2  + 
 \left [ \frac{Y_A - Y_B}{0.005} \right ]^2 +  \left [ \frac{Z_A -
 Z_B}{0.0005} \right ]^2. 
 \end{equation}
In this metric, $\tau_A$, $Y_A$ and $Z_A$  are the age, input helium abundance, and input heavy element mass fraction of the $\alpha$ Cen A model, {and $\tau_B$,  $Y_B$, and $Z_B$ are the same for the $\alpha$ Cen B model}. The denominators in the above equations represent a theoretical ``uncertainty''  which corresponds to half the grid resolutions of each variable. This permits small differences (e.g.\ within roughly $2\sigma$) in age or initial composition between models of $\alpha$ Cen A and~B.

\citet{GoodKroup} note that the fragmentation of a star-forming core into a binary or multi--star system occurs on time scales of less than 0.1 Myr, which suggests that paired models of $\alpha$ Cen A and B should differ in age by less than 0.1 Myr. However, our models do not take such small steps on the main sequence, and so are unlikely to have ages within this proximity. Our time steps along the main sequence are of order 10 Myr, and so we require that the age discrepancy between $\alpha$ Cen A and B is normalized by 5 Myr, representing agreement at the $2\sigma$ level. 


To determine the best-fitting model to non-seismic observations globally,
we consider weighted and unweighted combinations of the binary and
classical goodness-of-fit scores: 
\begin{equation*}
\begin{aligned}
s^2_{\text{classical,weighted}} &= \tfrac{1}{10}s^2_{\text{classical}}+
\tfrac{1}{6}s^2_{\text{binary}}%
\\
s^2_{\text{classical,unweighted}} &= s^2_{\text{classical}} +
s^2_{\text{binary}}. 
\end{aligned}
\end{equation*} 

We note that $s^2_{\text{weighted}}$ (hereafter $s^2_{\rmw}$) is similar to a reduced $\chi^2$ in which we have treated each of the components from the classical and binary constraints equally. The score $s^2_{\text{unweighted}}$ (hereafter $s^2_{\rmuw}$), on the other hand, is similar to a $\chi^2$ score with 8 degrees of freedom, 5 of which come from the independent stellar constraints and 3 of which come from the binary constraints. In this latter statistic, consistency with individual stellar properties is prioritized. We compute these $\chi^2$ minimized-style scores for each grid model.

Comparing between configurations, we find that the best classical grid scores, on average, are attained with models using Eddington boundary conditions (BCs). Among variations in diffusion, models with standard diffusion $\eta_D=1.0$ perform best, followed by models with suppressed and enhanced diffusion ($\eta_D=0.5$ and $\eta_D=1.5$), respectively, and then by enhanced-diffusion models which use convective overshoot. We note, however, that models with suppressed diffusion constitute the largest proportion of the well-fitting population. The models using Krishna Swamy BCs perform noticeably worse than others in terms of classical parameters. Table~\ref{cl} gives a basic statistical summary of the $3\sigma$ population, i.e.\ those models which fit better than $s^2_{\text{classical,w}}=3.0$. This set includes 1218 of the ${\sim}150,000$ generated tracks.

\begin{table}
\centering
\caption{Properties of Grid Models with $s^2_{\text{cl,w}}\le 3.0$}
\begin{tabular}{l cc cc}
\hline\hline
{Config} &
Number &

{$\alpha_{\text{A}} / \alpha_{\odot}$} &
{$\alpha_{\text{B}} / \alpha_{\odot}$} &

$s^2_{\text{cl,w}}$
\\ \hline
Standard		&  300	&  0.897  &  1.067  &  1.965  \\
KS			&  98  &  0.809  &  0.958  &  2.298  \\
Low Diffusion	&  422	&  0.900  &  1.053  &  2.003  \\
High Diffusion	&  252	&  0.891  &  1.066  &  2.047  \\
Overshoot		&  146	&  0.888  &  1.065  &  2.146  \\
\hline 
\end{tabular}
\tablecomments{ 
Mixing lengths are solar-normalized by configuration, according to Table
\ref{solar}. Mixing lengths and classical scores are averaged over the
number of models with the given configuration.
}
\label{cl}
\end{table}

\begin{figure}
\centering
\includegraphics[width=\columnwidth]{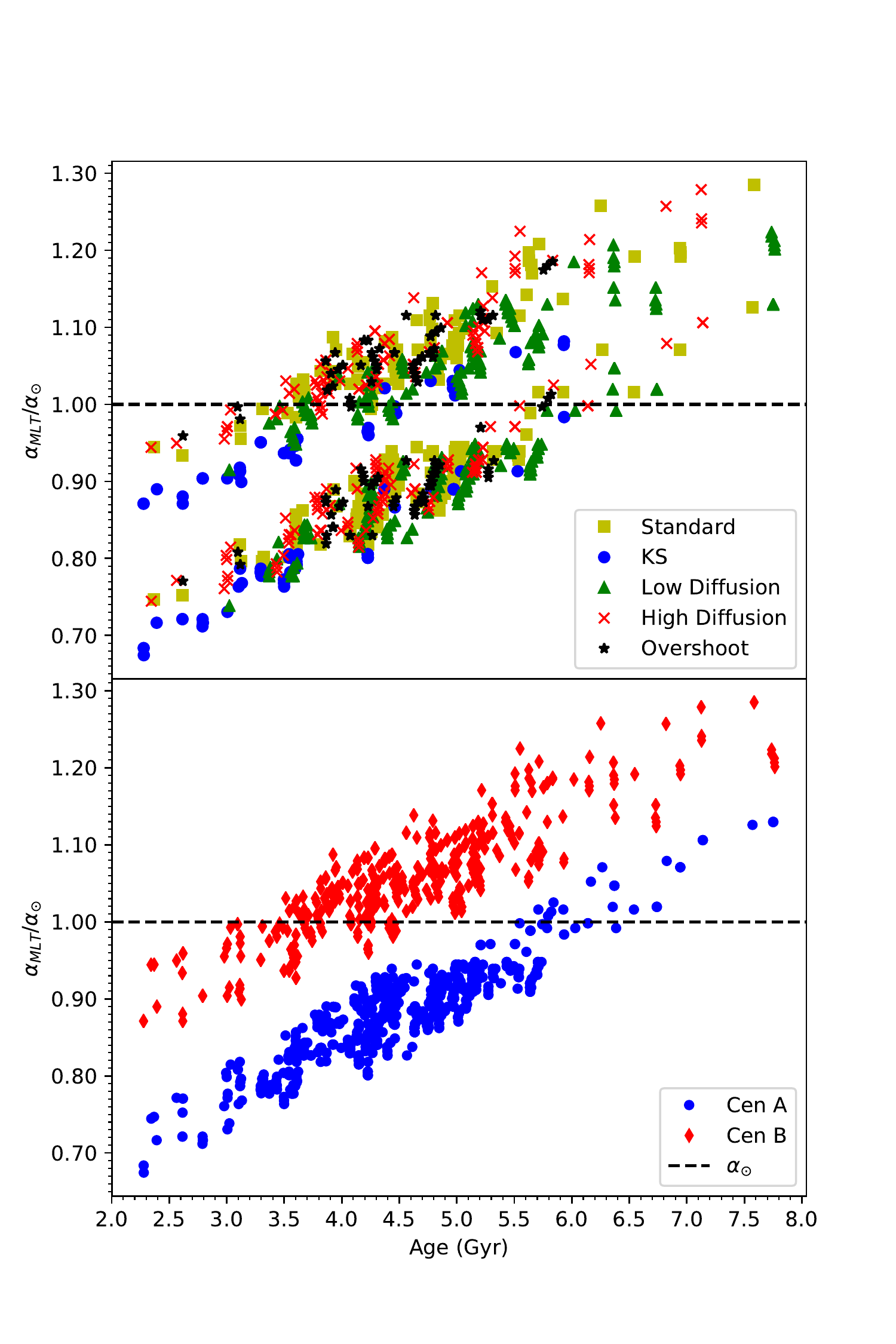} 
\caption{
Normalized mixing length is shown as a function of age for the $3\sigma$
classical model population. Solar mixing length is indicated with a
horizontal dashed line.  Top: Models are sorted according to physical
configuration, indicated by marker color and shape. Bottom: Models are
separated by star.
}
\label{mlt_vs_age_classic}
\end{figure}

Figure \ref{mlt_vs_age_classic} shows the $3\sigma$ sample of 1218 tracks as a function of age, sorted by configuration and by star. This and Table \ref{cl} suggest that the optimal mixing length for $\alpha$ Cen~A in a given model pair should be lower than the optimal mixing length for $\alpha$ Cen B. We find this to be true in strict terms; no fitting model in any configuration has $\alpha_{A,\text{MLT}} > \alpha_{B,\text{MLT}}$. This result is consistent with \citet{Joyce2018}, who found that stellar models tailored to fit the observed radius and luminosity of metal-poor sub-dwarf HD 140283 required lower mixing lengths for higher input masses. We also find that, for the vast majority of models, optimized mixing lengths for $\alpha$ Cen A are slightly sub-solar, whereas for B, they are super-solar.

\section{Asteroseismic Models}
\label{astero}

Though we find consistent classical models at the computational resolutions discussed in \S \ref{modproc}, helioseismic calculations \citep[e.g.][]{Schunker} have shown that, in order to obtain accurate non-radial oscillation frequencies, one must solve the stellar structure equations with a higher tolerance and use a more sophisticated equation of state. However, models with order-of-magnitude improvements in tolerances and which use a more sophisticated equation of state run at least an order of magnitude more slowly. 

Rather than run a grid of ${\sim} 150,000$ tracks with asteroseismic tolerances outright, we create a significantly reduced grid of models selected according to the classical optimization. We run an asteroseismic counterpart for each grid model in {an $\alpha$ Cen A, $\alpha$ Cen B} pair with $s^2_{\text{cl,w}} < 3.0$.  These models use tolerances of one part in $10^6$ and the FreeEOS \citep{FreeEOS} equation of state. Asteroseismic DSEP models generate structure files compatible with the GYRE \citep{GYRE} stellar oscillation code, which is then used to determine the resonant acoustic pressure modes. We verify that the seismic models produce the same classical results as their grid-quality counterparts. We find that the exact values for e.g.\ luminosity and radius differ at a level well below the grid resolution, and so we consider this method acceptable. This two-step procedure allows us to consider a cross-section of seismic data which is more likely to produce viable fits globally without extending computation time into weeks. 

A pair of {$\alpha$ Cen A and B} tracks  which fit the classical constraints is specified by 8 parameters: mass, mixing length, initial helium abundance, and the initial mass fraction of heavy elements, for each star. 
In total, there are 196 unique sets of input parameters which produce  $s^2_{\text{cl,w}} < 3.0$. This corresponds to a maximum of 392 individual models, but we note that the complete set may include duplicates of parameter combinations for an individual star. Asteroseismic-quality models are calculated with DSEP for each of 196 unique input parameter sets. Of these, 51 invoke standard physics, 16 use Krishna Swamy (KS) atmospheric BCs, 65 use low (suppressed) diffusion, 39 use high (enhanced) diffusion, and 15 use high diffusion and allow for convective overshoot in {$\alpha$ Cen A.} For the sake of uniformity in comparison, we also generate a seismic solarmodel for each physical configuration using the mixing lengths in Table \ref{solar}. Seismic data in Table \ref{solar} comes from these models.

\begin{figure}
\includegraphics[width=\linewidth]{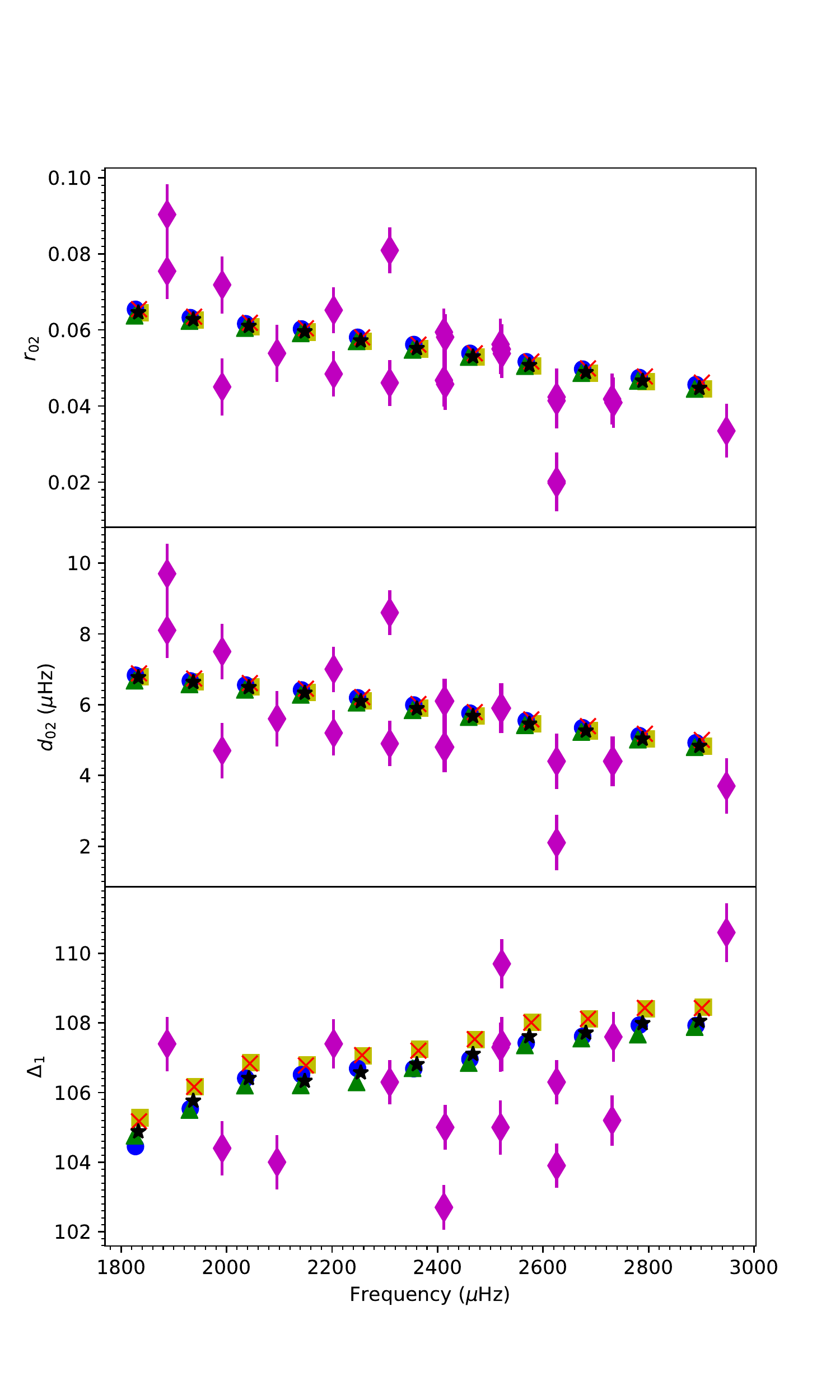} 
\caption{Theoretical and observed seismic parameters $\Delta_1$ (top), $d_{02}$ (middle), and $r_{02}$ (bottom) are shown for $\alpha$ Cen A as a function of frequency $\nu(n)$ for radial orders  $n=16$--$26$. Observations are shown with error bars in pink. Seismic parameters for the best-fitting model of each physical configuration are shown.  Marker convention is the same as the legend in Figure \ref{mlt_vs_age_classic}. Theoretical values among different configurations are largely overlapping. }
\label{rA}
\end{figure}

\begin{figure}
\includegraphics[width=\linewidth]{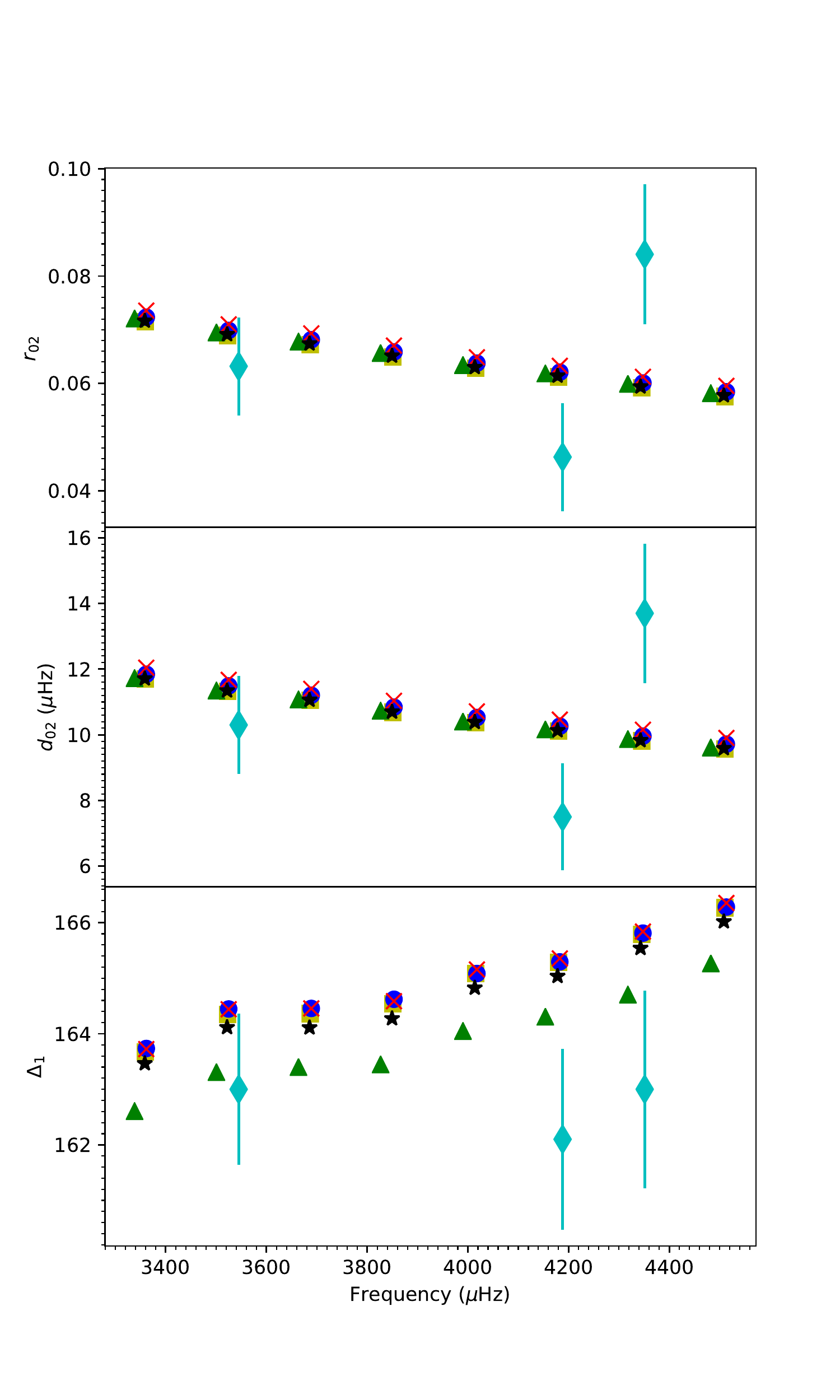}
\caption{
The same as Figure \ref{rA} but for $\alpha$ Cen B, showing radial orders
$n=19$--$26$ and observations in blue with error bars. Theoretical values
are designated as in Figure \ref{rA} and are again largely overlapping.}
\label{rB}
\end{figure}

The observations of {$\alpha$ Cen A and $\alpha$ Cen B} are not complete across all modes. Since a single calculation of $r_{02}(n)$ requires measurements of four particular frequencies, each with a different harmonic signature, the number of individual $r_{0,2}$ values available from the observational data for $\alpha$ Cen is not large. 
The theoretical ratios, on the other hand, do not suffer from any incompleteness.  This is illustrated in Figures \ref{rA} and \ref{rB}, which show theoretical~$r_{02}$ (and $\Delta_1, d_{02}$) values as a function of frequency for the best-fitting seismic model of each configuration compared to the observational data.  The sparseness in the $\alpha$ Cen A and B data is evident here, which supports using an average $r_{02}$ value when comparing the models to the observations.

\section{Comparison to Seismic Observations}
\label{math}
Once we obtain theoretical $p$ modes from GYRE for both seismic tracks in an $\alpha$ Cen model, we recover a goodness-of-fit between the theoretical and observed values of the average ratio of the small and large frequency separations, $r_{02}$, as defined in \S \ref{obs}. The weighted seismic agreement score is given by
\begin{equation}
s^2_{\text{seismic,w}}  =   \frac{1}{2} \left [ \frac{r_{\text{A,obs}}
- r_{\text{A,mod}}}{\sigma_{r_{02, \text{A}} }} \right]^2  +
\left [ \frac{r_{\text{B,obs}} - r_{\text{B,mod}}}{\sigma_{r_{02,
\text{B}}} } \right ]^2, 
\label{seis_score}
\end{equation}
where $r_{\text{A,obs}}$ and $\sigma_{r_{0,2},\text{A}}$ are the average ratio and uncertainty taken directly from \citet{deMeulenaer}, and $r_{\text{B,obs}}$ and $\sigma_{r_{0,2},\text{B}}$ are taken directly from \citet{Kjeldsen}. 

In calculating $s^2_{\text{seismic,w}}$, we first homogenize the observed and synthetic data by including frequency contributions to theoretical $r_{02}$ only for the radial orders for which there are observations. For $\alpha$ Cen A, this is $n=16$--$26$, and for $\alpha$ Cen B, $n=19$--$26$. Table \ref{syspar} gives observed values and uncertainties for seismic parameters $\Delta_1$, $ d_{02}$, and $r_{02}$.

While other authors have adopted more complicated statistics to quantify theoretical--observational consistency for seismic models, the literature suggests that using average $r_{02}$ yields nearly the same conclusions. In fact, \citet{Aguirre13} and \citet{Lebreton} find that use of the frequency ratios, rather than e.g.\ individual frequencies or scaling relations, produces more precise asteroseismic ages. Since we have well-determined values of average $r_{02}$ from the observers directly, we use this method.

\subsection{Seismic Constraints on Age}

Among 192 unique pairs of tracks, 27 $\alpha$ Cen models in total are found to produce seismic agreement within~$3\sigma$ (i.e.\ $s^2_{\text{seismic,w}}\le3$). This includes at least one model with each set of input physics. Among the 27-model seismic population, 8 use standard physics, one uses KS BCs, 12 have low diffusion, three have high diffusion, and three have high diffusion and overshoot in {$\alpha$ Cen A.} Among the seismically fitting models, only two models have a convective core in $\alpha$ Cen A, and both use high diffusion and overshoot.

\begin{figure}
\includegraphics[width=\linewidth]{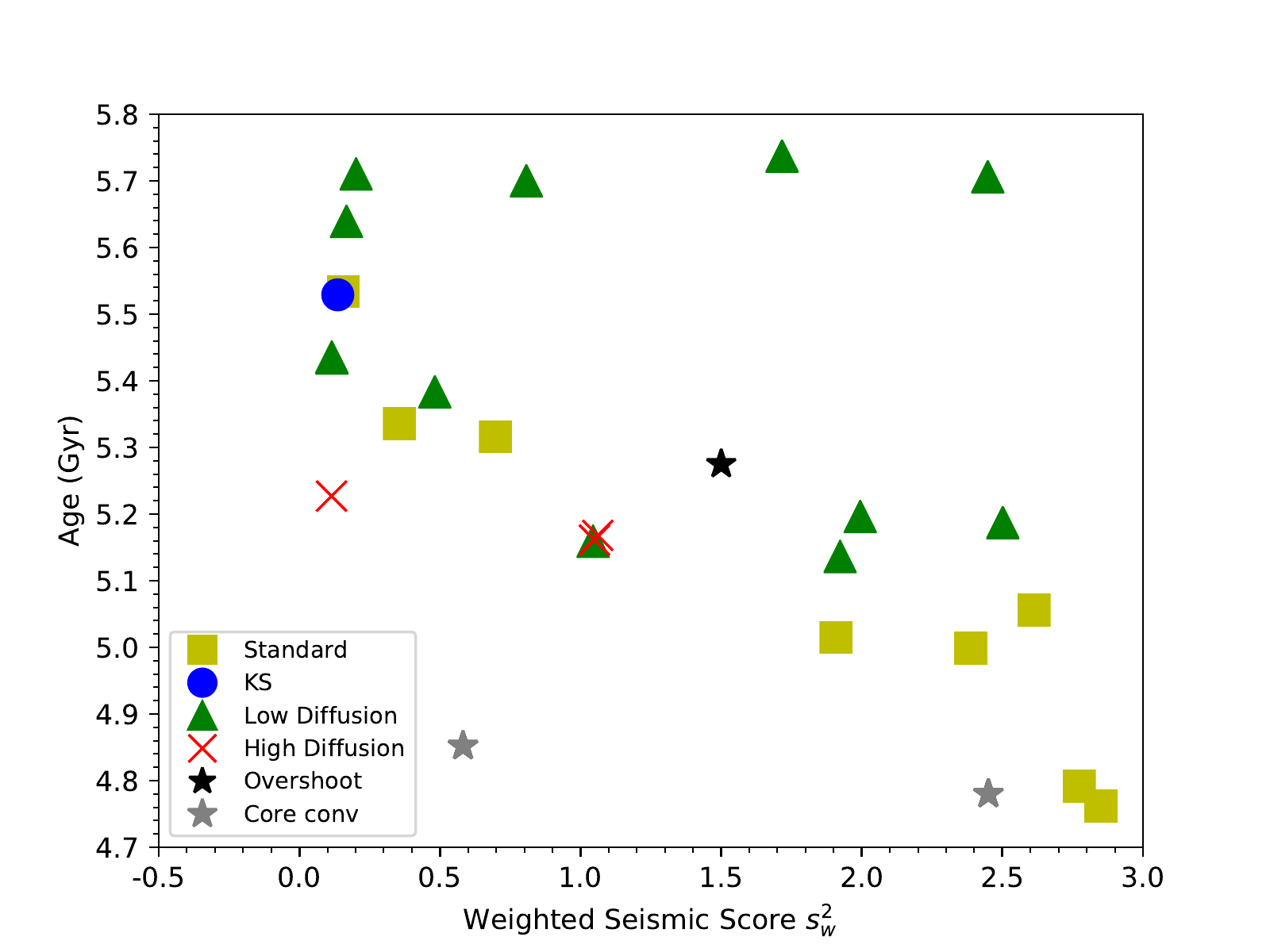}
\caption{
Model ages are shown as a function of $s^2_{\text{s,w}}$. Models with $3
\sigma$ or better seismic agreement are shown. Overshoot models (stars)
are separated into those which find core convection in Cen~A (gray)
and those which do not (black). 
}
\label{config_age}
\end{figure}

The overall distribution of goodness-of-fit by physical prescription suggests a preference among well-fitting seismic models for two features in particular. First, there is a preference for Eddington surface boundary conditions. Just under 30\% of the sub-$s^2_{\text{s,w}}=3$ sample is composed of models which use Eddington BCs and standard physics, versus 4\% with Krishna Swamy BCs and standard physics. If the seismic sample is extended to consider all models with $s^2_{\text{s,w}} \le 6.0$ (resulting in 42 pairs), these percentages remain the same. 
Second, there is a strong preference for models with suppressed diffusion---these make up nearly half of the $s^2_{\text{seismic,w}} \le 3.0$ sample. It is also found that models with $\eta_D = 0.5$ constitute a greater proportion of the sample as we restrict to tighter seismic agreement scores. Low diffusion is used in 7 of the 15 best-fitting seismic models, 6 of the top 12, and 5 of the top 10.  Figure \ref{config_age} shows age as a function of seismic score for the $\le 3\sigma$ model set, where the preference for standard boundary conditions and low diffusion is evident. 

We note that the atmospheric prescription does not affect $r_{02}$ (which is designed so that it is sensitive only to the interior of the star), but that surface boundary choices do affect the classical (e.g.\ surface) observables. Models with Krishna Swamy surface boundary conditions which fit the classical observations have interior structures which are rarely compatible with the seismic observations.

Since seismic parameters characterize the stellar interior, comparing between modeling prescriptions which vary the interior physics (e.g.\ diffusion and convective overshoot) is most relevant. Among three options for diffusive efficiency (without convective overshoot), there is a clear ordering: low diffusion does best, followed by standard diffusion, and lastly by high diffusion. Between high-diffusion models which allow for core overshoot and those that do not, models without convective overshoot fare somewhat better. There is no clear difference in observational agreement between convective core models (of {$\alpha$ Cen A)} and non-convective core models, but there is a large difference in fitted age.

There is a general trend of better agreement with older age among the $3\sigma$ seismic population, but this is most pronounced for standard models. Low-diffusion models lead to noticeably higher ages for the system than the others, and high diffusion leads to fitted ages which are lower by roughly 1 Gyr, on average, than fitted ages from models with low diffusion. The single, well-fitting {$\alpha$ Cen A} model with KS BCs is isolated. The two models which find convective cores in $\alpha$ Cen A---both high-diffusion models with overshoot---have considerably lower fitted ages than those found by the rest of the models.

As we noted in \citet{Joyce2018}, a model with lower diffusive efficiency will have increased hydrogen abundance in the core, resulting in a longer hydrogen burning phase (i.e.\ longer main sequence lifetime), slower evolution, and hence older intersection age. This means that, for a fixed point in time on the HR diagram, a low-$\eta_D$ model will have a less evolved interior structure than a model with higher $\eta_D$ (with all other features being the same).
Since $r_{02}$ is indicative of age, the fact that low-diffusion models produce compatible interior structures at older ages than standard models, which in turn find older ages than high-diffusion models, is consistent with \citet{Joyce2018}.  Further, we can consider differences in overshoot models which do or do not find a convective core in $\alpha$ Cen A. While all overshoot models produce ages lower than the bulk of the population by 0.5 to 1.0 Gyr, fitted ages for the subsample with core convection are an additional ${\sim}0.5$ Gyr lower than the model without. We elaborate more on core properties and age in Section \ref{results}.

\subsection{Seismic Constraints on Mixing Length}

\begin{figure}
\centering
\includegraphics[width=\columnwidth]{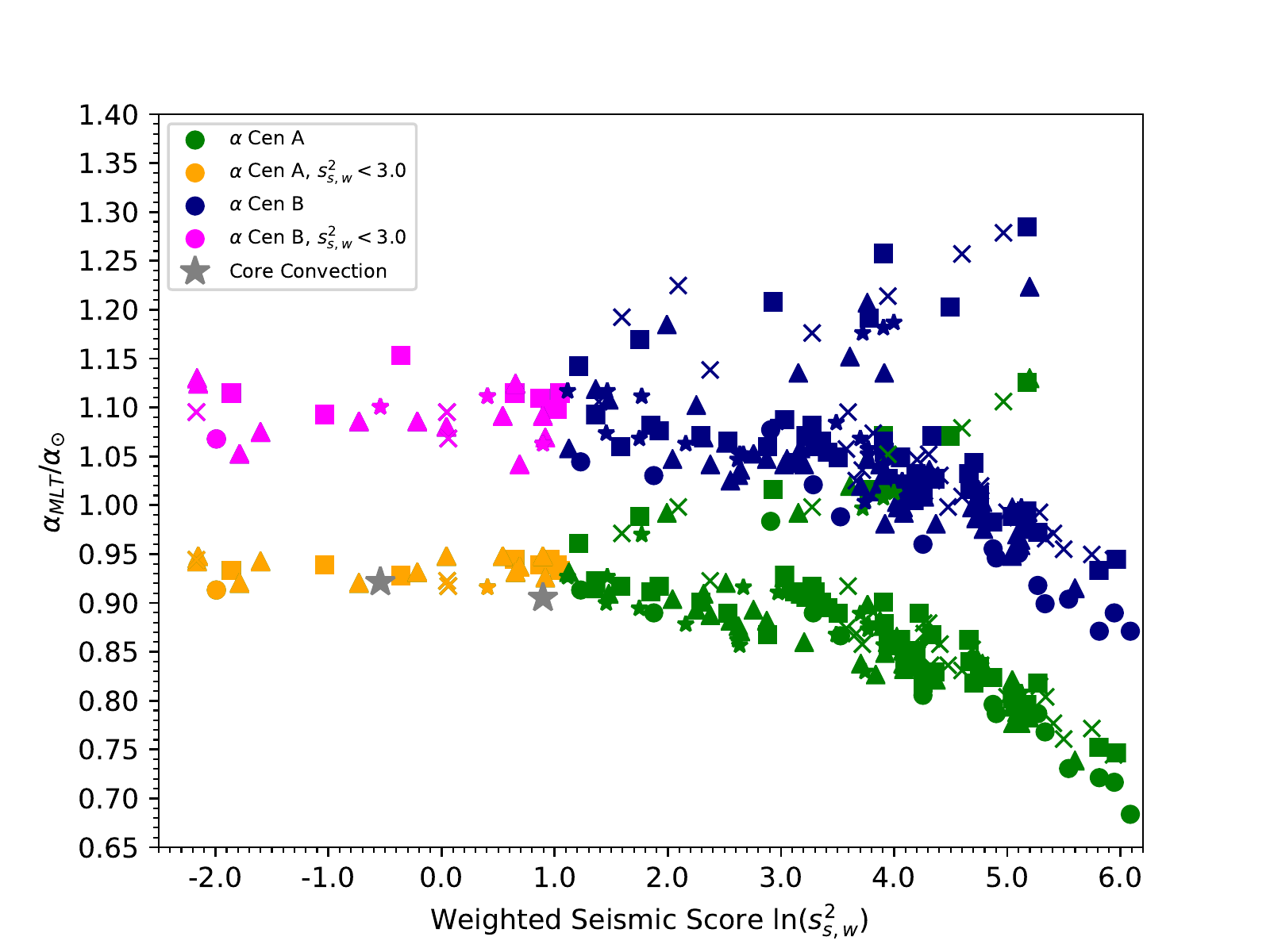} 
\caption{Models producing $s^2_{\text{s,w}} \le 3$ are highlighted. {$\alpha$ Cen A} models with convective cores are indicated with gray stars. Configurations are indicated by the same marker style conventions as in previous figures.} 
\label{mlt_vs_seismic}
\end{figure}

Figure \ref{mlt_vs_seismic} shows mixing length as a function of seismic score for {$\alpha$ Cen A and B} across all 192 pairs of seismic models. Seismic scores range from ${\sim}0.1$ to 440, so are presented in terms of $\ln (s^2_{\text{s,w}})$.

Interestingly, mixing lengths which are optimal according to seismic score converge fairly closely to the same (normalized) value regardless of choices in input physics, especially for {$\alpha$ Cen A.} We also see clearly that seismic models of {$\alpha$ Cen A} are more sensitive to the choice of $\alpha_{\text{MLT,A}}$ than are models of {$\alpha$ Cen B} to $\alpha_{\text{MLT,B}}$. This is evident in the tighter vertical clustering of the {$\alpha$ Cen A} data across the entire score regime, but especially among models with low $s^2_{\text{s,w}}$.

The observed frequencies constrain the allowed interior structure of the models, which explains the pronounced sensitivity to mixing length when seismic constraints are included.  Since {$\alpha$ Cen A} has a higher mass than {$\alpha$ Cen B}, it evolves more rapidly, which means the variation in interior structure for {$\alpha$ Cen A} models is greater. It is thus not surprising that the choice of mixing length parameter affects {$\alpha$ Cen A more than $\alpha$ Cen B.}

A global measure of agreement between the models and all of the observations is obtained by combining the classical and binary constraints with the seismic criterion.  We compute weighted and unweighted forms:
\begin{align}
s^2_{\text{total,w}} &= \tfrac{1}{2} ( s^2_{\text{cl, w}}+
s^2_{\text{seismic, w}} ), \\
s^2_{\text{total}} &= s^2_{\text{classical}} + s^2_{\text{seismic}} \, .
\end{align}  
Here, $s^2_{\text{total, w}}$  is analogous to a reduced $\chi^2$ score with equal weighting among the classical, seismic, and binary constraints, and $s^2_{\text{total,unweighted}}$ is similar to a $\chi^2$ score with 10 degrees of freedom, where 5 degrees come from classical observations, 2 degrees from the seismic observations, and 3 degrees from the binary constraints. Among the complete set of seismic models, we find seismic scores as high as $s^2_{\text{s,w}}=400$, and the vast majority exceed $s^2_{\text{s,w}}=3$. As such, the seismic agreement statistic heavily dominates the global scores.

We find 31 models which are consistent at the level $s^2_{\text{tot,w}} \le 3$, a slight expansion over the set of 27 seismically viable pairs. Table \ref{final} shows the classical, seismic, and total weighted scores for this set, as well as key classical and seismic parameters.

\begin{table*} 
\centering 
\caption{Key Parameters for Best-fitting Models}
\begin{tabular}{c ll c lll lll	lll }
\hline\hline
Rank &

Star &
Config &
$\alpha /\alpha_{\odot}$ &

$Z_{\text{in}}$ &
$Y_{\text{in}}$ &
Age &

$\Delta_{1,\text{th}}$ &
$d_{0,2,\text{th}}$ &
$r_{0,2,\text{th}}$ &

$s^2_{\text{cl,w}}$ &
$s^2_{\text{seis,w}}$  &
$s^2_{\text{tot,w}}$ 
\\ \hline
1&  {$\alpha$ Cen A}$^{\dag}$ & Standard &  0.934	&  0.024  &  0.26  &  5.53  &
107.3  &  5.9  &  0.055  &  1.56  &  0.16  &  0.86  \\ 
 & {$\alpha$ Cen B}$^{\dag}$ &  &	1.115  &  0.025  &  0.26  &  5.54  &  164.9  &
 10.6  &  0.064  &  1.56  &  0.16  &  0.86  \\
2&  {$\alpha$ Cen A}		& Standard &  0.939  &	0.025  &  0.27	&  5.34
&
107.0  &  5.8  &  0.054  &  1.86  &  0.36  &  1.11  \\ 
 & {$\alpha$ Cen B}		&  &  1.093  &	0.025  &  0.26	&  5.35  &
 164.9	&  10.7  &  0.065  &  1.86  &  0.36  &	1.11  \\
3&  {$\alpha$ Cen A}$^{\dag}$ & Low Diff &  0.920	&  0.027  &  0.28  &  5.38  &
106.7  &  5.9  &  0.056  &  1.79  &  0.48  &  1.14  \\ 
 & {$\alpha$ Cen B}$^{\dag}$ & &  1.086  &  0.027	&  0.27  &  5.37  &  165.0  &
 10.7  &  0.065  &  1.79  &  0.48  &  1.14  \\
4&  {$\alpha$ Cen A}		& Low Diff &  0.948  &	0.029  &  0.28	&  5.43
&
107.2  &  5.9  &  0.055  &  2.30  &  0.12  &  1.21  \\ 
 & {$\alpha$ Cen B}		&  &  1.124  &	0.03  &  0.27  &  5.45	&  165.3
 &  10.7  &  0.064  &  2.30  &	0.12  &  1.21  \\
5&  {$\alpha$ Cen A}		& Low Diff &  0.942  &	0.029  &  0.28	&  5.43
&
106.9  &  5.9  &  0.055  &  2.42  &  0.12  &  1.27  \\ 
 & {$\alpha$ Cen B}		&  &  1.130  &	0.03  &  0.27  &  5.42	&  165.6
 &  10.7  &  0.065  &  2.42  &	0.12  &  1.27  \\
6&  {$\alpha$ Cen A}$^{\dag}$ & High Diff &  0.944  &  0.028  &  0.28  &  5.23  &
106.6  &  5.8  &  0.055  &  2.44  &  0.11  &  1.28  \\ 
 & {$\alpha$ Cen B}$^{\dag}$ &  &	1.095  &  0.028  &  0.27  &  5.24  &  163.9  &
 10.6  &  0.065  &  2.44  &  0.11  &  1.28  \\
7&  {$\alpha$ Cen A}		& Standard &  0.928  &	0.026  &  0.27	&  5.32
&
106.8  &  6.0  &  0.056  &  2.07  &  0.70  &  1.38  \\ 
 & {$\alpha$ Cen B}		&  &  1.153  &	0.027  &  0.26	&  5.31  &
 165.8	&  10.7  &  0.065  &  2.07  &  0.70  &	1.38  \\
8&  {$\alpha$ Cen A}$^{\dag \star }$ & Overshoot &  0.921	&  0.026  &  0.28  &
4.85  &  107.2	&  6.0	&  0.056  &  2.29  &  0.58  &  1.44  \\ 
 & {$\alpha$ Cen B}$^{\dag}$ &  &	1.101  &  0.027  &  0.27  &  4.86  &  165.0  &
 10.9  &  0.066  &  2.29  &  0.58  &  1.44  \\
9&  {$\alpha$ Cen A} & Low Diff &	0.920  &  0.025  &  0.27  &  5.64  &  106.8  &
5.8  &	0.054  &  2.75	&  0.17  &  1.46  \\ 
 & {$\alpha$ Cen B} &  &  1.052  &  0.025	&  0.26  &  5.62  &  164.0  &  10.6  &
 0.065	&  2.75  &  0.17  &  1.46  \\
10&  {$\alpha$ Cen A}$^{\dag}$ & KS &  0.913  &  0.024  &	0.26  &  5.53  &  106.8
&
5.9  &	0.055  &  2.86	&  0.14  &  1.50  \\ 
 & {$\alpha$ Cen B}$^{\dag}$ &  &	1.068  &  0.025  &  0.26  &  5.51  &  164.7  &
 10.6  &  0.064  &  2.86  &  0.14  &  1.50  \\
11&  {$\alpha$ Cen A}  & Low Diff &  0.942	&  0.026  &  0.27  &  5.71  &  107.4  &
5.9  &	0.055  &  2.97	&  0.20  &  1.59  \\ 
 & {$\alpha$ Cen B}  &  &  1.074  &  0.028	&  0.27  &  5.71  &  164.0  &  10.5  &
 0.064	&  2.97  &  0.20  &  1.59  \\
12&  {$\alpha$ Cen A}  & Overshoot &  0.916  &  0.024  &  0.27  &  5.28  &	106.9  &
5.7  &	0.053  &  1.83	&  1.50  &  1.67  \\ 
 & {$\alpha$ Cen B}  & &  1.111  &	0.025  &  0.26	&  5.27  &  164.9  &  10.7  &
 0.065	&  1.83  &  1.50  &  1.67  \\
13&  {$\alpha$ Cen A}  & Low Diff &  0.948	&  0.03  &  0.29  &  5.16  &  107.2  &
6.0  &	0.056  &  2.36	&  1.04  &  1.70  \\ 
 & {$\alpha$ Cen B}  & &  1.080  &	0.03  &  0.28  &  5.14	&  164.9  &  10.8  &
 0.066	&  2.36  &  1.04  &  1.70  \\
14&  {$\alpha$ Cen A}  & High Diff &  0.917  &  0.025  &  0.27  &  5.17  &	106.9  &
6.0  &	0.056  &  2.37	&  1.06  &  1.71  \\ 
 & {$\alpha$ Cen B}  & &  1.068  &	0.025  &  0.26	&  5.18  &  163.9  &  10.7  &
 0.066	&  2.37  &  1.06  &  1.71  \\
15&  {$\alpha$ Cen A}  & Standard &  0.945	&  0.029  &  0.29  &  5.01  &  107.0  &
6.1  &	0.057  &  1.88	&  1.91  &  1.89  \\ 
 & {$\alpha$ Cen B}  &  &  1.115  &  0.03  &  0.28	&  5.02  &  165.0  &  10.8  &
 0.066	&  1.88  &  1.91  &  1.89  \\
16&  {$\alpha$ Cen A}  & Low Diff &  0.931	&  0.026  &  0.27  &  5.70  &  106.7  &
5.7  &	0.054  &  3.00	&  0.81  &  1.90  \\ 
 & {$\alpha$ Cen B}  &  &  1.086  &  0.028	&  0.27  &  5.70  &  164.5  &  10.5  &
 0.064	&  3.00  &  0.81  &  1.90  \\
17&  {$\alpha$ Cen A}  & High Diff &  0.923  &  0.025  &  0.27  &  5.16  &	107.2  &
6.0  &	0.056  &  2.81	&  1.05  &  1.93  \\ 
 & {$\alpha$ Cen B}  &  &  1.095  &  0.025	&  0.26  &  5.14  &  164.9  &  10.8  &
 0.065	&  2.81  &  1.05  &  1.93  \\
18&  {$\alpha$ Cen A}  & Low Diff &  0.948	&  0.026  &  0.27  &  5.74  &  107.2  &
5.7  &	0.053  &  2.39	&  1.72  &  2.05  \\ 
 & {$\alpha$ Cen B}  &  &  1.091  &  0.025	&  0.26  &  5.75  &  164.9  &  10.5  &
 0.064	&  2.39  &  1.72  &  2.05  \\
19&  {$\alpha$ Cen A}  & Standard &  0.961	&  0.028  &  0.28  &  5.61  &  106.5  &
5.6  &	0.052  &  0.93	&  3.36  &  2.14  \\ 
 & {$\alpha$ Cen B}  &  &  1.142  &  0.028	&  0.27  &  5.61  &  165.0  &  10.5  &
 0.063	&  0.93  &  3.36  &  2.14  \\
20&  {$\alpha$ Cen A}  & Low Diff &  0.937	&  0.028  &  0.28  &  5.20  &  107.4  &
6.1  &	0.057  &  2.31	&  1.99  &  2.15  \\ 
 & {$\alpha$ Cen B}  &  &  1.041  &  0.027	&  0.27  &  5.21  &  164.0  &  10.8  &
 0.066	&  2.31  &  1.99  &  2.15  \\
21&  {$\alpha$ Cen A}  & Standard &  0.939	&  0.03  &  0.29  &  5.00  &  106.6  &
6.1  &	0.057  &  2.21	&  2.39  &  2.30  \\ 
 & {$\alpha$ Cen B}  &  &  1.109  &  0.03  &  0.28	&  4.98  &  165.0  &  10.9  &
 0.066	&  2.21  &  2.39  &  2.30  \\
22&  {$\alpha$ Cen A}  & Low Diff &  0.931	&  0.029  &  0.29  &  5.14  &  106.9  &
6.1  &	0.057  &  3.00	&  1.92  &  2.46  \\ 
 & {$\alpha$ Cen B}  &  &  1.124  &  0.029	&  0.27  &  5.14  &  165.8  &  10.9  &
 0.066	&  3.00  &  1.92  &  2.46  \\
23&  {$\alpha$ Cen A}  & Standard &  0.945	&  0.028  &  0.28  &  5.06  &  107.4  &
6.1  &	0.057  &  2.39	&  2.61  &  2.50  \\ 
 & {$\alpha$ Cen B}  &  &  1.098  &  0.027	&  0.27  &  5.07  &  165.0  &  10.9  &
 0.066	&  2.39  &  2.61  &  2.50  \\
24 &  {$\alpha$ Cen A}$^{\star}$  & Overshoot &  0.927  &  0.024  &  0.27  &  5.32	&
107.0  &  5.6  &  0.052  &  2.12  &  3.05  &  2.58  \\ 
 & {$\alpha$ Cen B}  &  &  1.117  &  0.025	&  0.26  &  5.31  &  165.0  &  10.6  &
 0.065	&  2.12  &  3.05  &  2.58  \\
25&  {$\alpha$ Cen A}  & Overshoot &  0.905  &  0.026  &  0.28  &  4.78  &	107.1  &
6.1  &	0.057  &  2.81	&  2.45  &  2.63  \\ 
 & {$\alpha$ Cen B}  &  &  1.063  &  0.027	&  0.27  &  4.79  &  164.0  &  11.0  &
 0.067	&  2.81  &  2.45  &  2.63  \\
26&  {$\alpha$ Cen A}  & Low Diff &  0.948	&  0.028  &  0.28  &  5.71  &  106.7  &
5.6  &	0.053  &  2.88	&  2.45  &  2.67  \\ 
 & {$\alpha$ Cen B}  &  &  1.091  &  0.028	&  0.27  &  5.69  &  164.7  &  10.5  &
 0.064	&  2.88  &  2.45  &  2.67  \\
27&  {$\alpha$ Cen A}  & Low Diff &  0.926	&  0.027  &  0.28  &  5.19  &  107.1  &
6.1  &	0.057  &  2.98	&  2.50  &  2.74  \\ 
 & {$\alpha$ Cen B}  &  &  1.069  &  0.027	&  0.27  &  5.21  &  165.0  &  10.9  &
 0.066	&  2.98  &  2.50  &  2.74  \\
28&  {$\alpha$ Cen A}  & High Diff &  0.917  &  0.026  &  0.28  &  4.92  &	106.9  &
6.2  &	0.058  &  1.59	&  4.02  &  2.80  \\ 
 & {$\alpha$ Cen B}  &  &  1.106  &  0.027	&  0.27  &  4.92  &  164.9  &  10.9  &
 0.066	&  1.59  &  4.02  &  2.80  \\
29&  {$\alpha$ Cen A}  & Standard &  0.939	&  0.029  &  0.29  &  4.79  &  107.0  &
6.1  &	0.057  &  2.95	&  2.77  &  2.86  \\ 
 & {$\alpha$ Cen B}  &  &  1.098  &  0.029	&  0.27  &  4.79  &  164.8  &  11.0  &
 0.067	&  2.95  &  2.77  &  2.86  \\
30&  {$\alpha$ Cen A}  & Standard &  0.934	&  0.029  &  0.29  &  4.76  &  106.7  &
6.1  &	0.057  &  2.98	&  2.85  &  2.92  \\ 
 & {$\alpha$ Cen B}  &  &  1.115  &  0.029	&  0.27  &  4.76  &  165.4  &  11.1  &
 0.067	&  2.98  &  2.85  &  2.92  \\
31&  {$\alpha$ Cen A}  & KS &  0.913  &  0.028  &  0.28  &	5.04  &  107.0	&  6.1	&
0.057  &  2.48	&  3.42  &  2.95  \\ 
 & {$\alpha$ Cen B}  &  &  1.044  &  0.027	&  0.27  &  5.03  &  164.8  &  10.9  &
 0.066	&  2.48  &  3.42  &  2.95  \\
\hline
\end{tabular}
\tablecomments{A summary of the critical parameters and recovery statistics is provided for all models with $3\sigma$ or better agreement according to both classical and asteroseismic constraints. The best-fitting model from each physical configuration is indicated with $\dag$. The $8^{\text{th}}$- and $24^{\text{th}}$-best models have core convection in {$\alpha$ Cen A}, indicated with $\star$.
}
\label{final}
\end{table*}

\section{Results}
\label{results}
\begin{figure}
\centering
\includegraphics[width=\columnwidth]{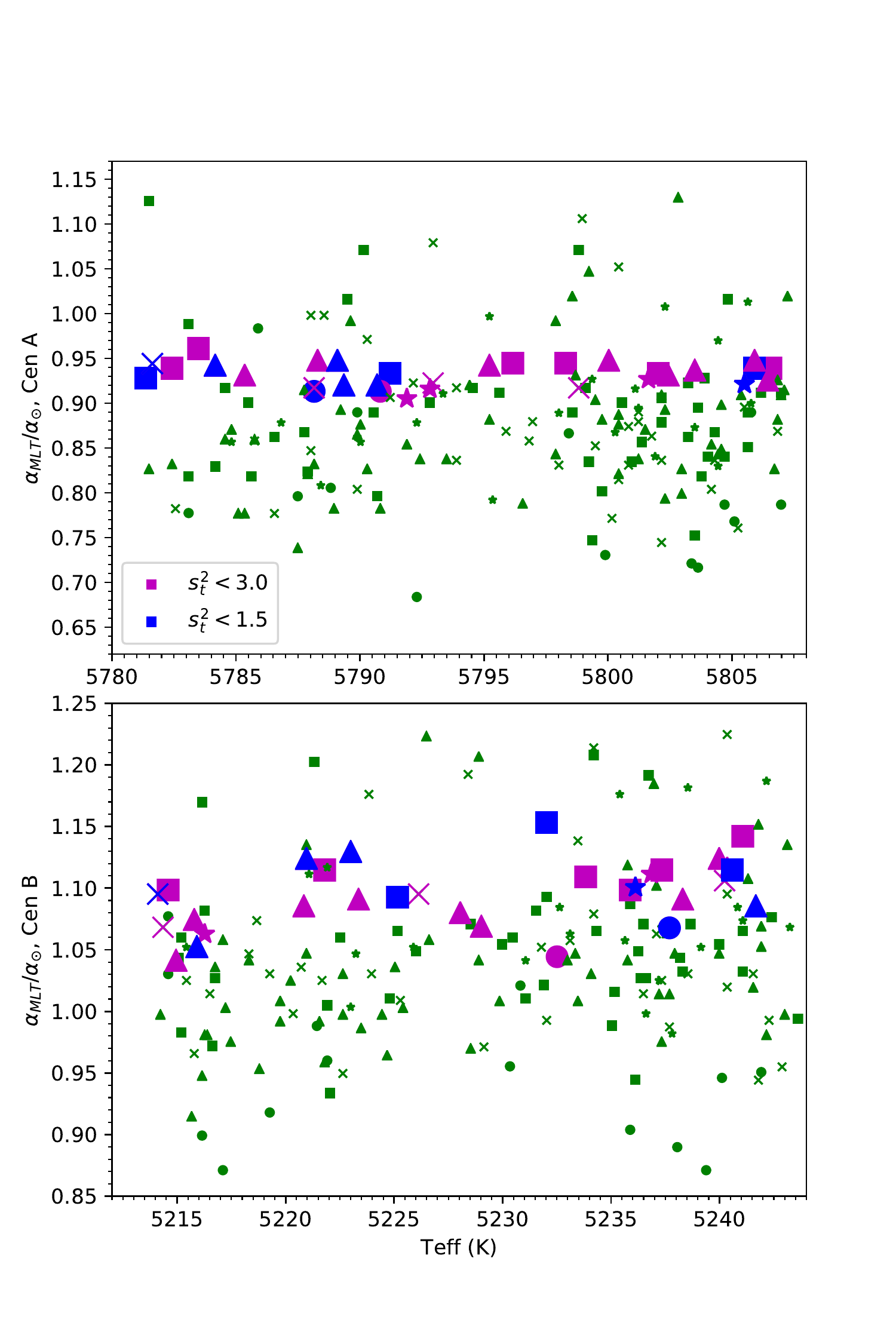}
\caption{
Viable mixing lengths are determined according to the total agreement
score and highlighted in purple ($s^2_{\text{tot,w}} \le3.0 $) and blue
($s^2_{\text{tot,w}} \le1.5 $) enlarged markers. Configurations are
indicated by marker style as in previous figures.
}	
\label{mlt_vs_temp}
\end{figure}

Figure \ref{mlt_vs_temp} presents the solar-normalized mixing lengths as a function of effective temperature for the 192 model sample with optimal values ($s^2_{\text{tot, w}}<3$ and $s^2_{\text{tot, w}}<1.5$) highlighted. We can see clearly that the optimized models prefer a particular mixing length for $\alpha$ Cen A regardless of the choice of input physics, converging to a value of $\alpha_{\text{MLT}}/\alpha_{\odot} = 0.932$ averaged over all models with $s^2_{\text{t,w}}\le3$.
We note that the introduction of seismic constraints significantly reduces the scatter in viable $\alpha_{\text{MLT,A}}$, and that the results are higher on average than the classically optimized mixing lengths. In short, the best fit for $\alpha_{\text{MLT,A}}$ is sub-solar, consistent across configurations, and heavily determined by the seismic constraints. This is not too surprising, as there is a clear correlation between age and mixing length in the models (see for instance Figure \ref{mlt_vs_age}), and the seismic observations provide information on the evolutionary state of the star. 

There is  more scatter among the optimal mixing lengths for {$\alpha$ Cen B}. The values are, however, super-solar regardless of the choice of input physics. There is a reduction in scatter between the $\alpha_{\text{MLT,B}}$ values allowed by classical constraints and those allowed globally, though less severely than for {$\alpha$Cen A.}

There is a clear relationship between mass and mixing length for three stars---{$\alpha$ Cen A, $\alpha$ Cen B}, and the Sun---which follows the same trend observed in \citet{Joyce2018} among fits to sub-dwarf HD 140283.	{$\alpha$ Cen A,} having the largest mass at $1.10$--$1.11\, M_{\odot}$, requires that $\alpha_{\text{MLT,A}}$ is less than $\alpha_{\odot}$ by between 6\% and  8\%. {$\alpha$ Cen B}, at $0.93$--$0.94\, M_{\odot}$, requires that $\alpha_{\text{MLT,A}} $ is greater than~$\alpha_{\odot}$ by 8\% to 11\%.  The trend is preserved across five different prescriptions for modeling physics, and mixing lengths for {$\alpha$ Cen A and $\alpha$ Cen B} best converge when all observational constraints are included in the agreement statistic. The mixing length choices are especially sensitive to seismic parameters, and there is almost no deviation in the optimal mixing length for $\alpha$ Cen A among 31 candidates.

In their Table 6, \citet{Viani} predict mixing lengths as a function of surface gravity ($\log g$), effective temperature, and metallicity (in terms of [Fe/H]) with fitting coefficients varying dependent on their physical prescriptions. Passing these parameters for {$\alpha$ Cen A and $\alpha$ Cen B} from our optimal models, we find that their predicted mixing lengths for {$\alpha$ Cen B}, $\alpha/\alpha_{\odot}\sim1.11$ on average, are in good agreement with ours, particularly when coefficients for their ``Non-diffusion, $0.2H_p$'' and ``No $\nu_{\text{max}}$'' prescriptions are used. The values they predict for {$\alpha$ Cen A}, however, are always super-solar, $\alpha/\alpha_{\odot}\sim1.065$ on average, and in fact are very similar to the values they predicted for {$\alpha$ Cen B.} In most cases, the mixing lengths they predict for our {$\alpha$ Cen A} parameters are slightly smaller than those for {$\alpha$ Cen B}, but there is no appreciable difference.

\citet{Creevey17} likewise provide a mixing length equation based on the same physical attributes as \citet{Viani}. For our 31 models, their equation (5) gives average mixing lengths of $\alpha_{\text{MLT,A}} / \alpha_{\odot}=0.962$ and $\alpha_{\text{MLT,B}} / \alpha_{\odot}=1.073$, both in excellent agreement with our findings.

\begin{figure}
\centering
\includegraphics[width=\columnwidth]{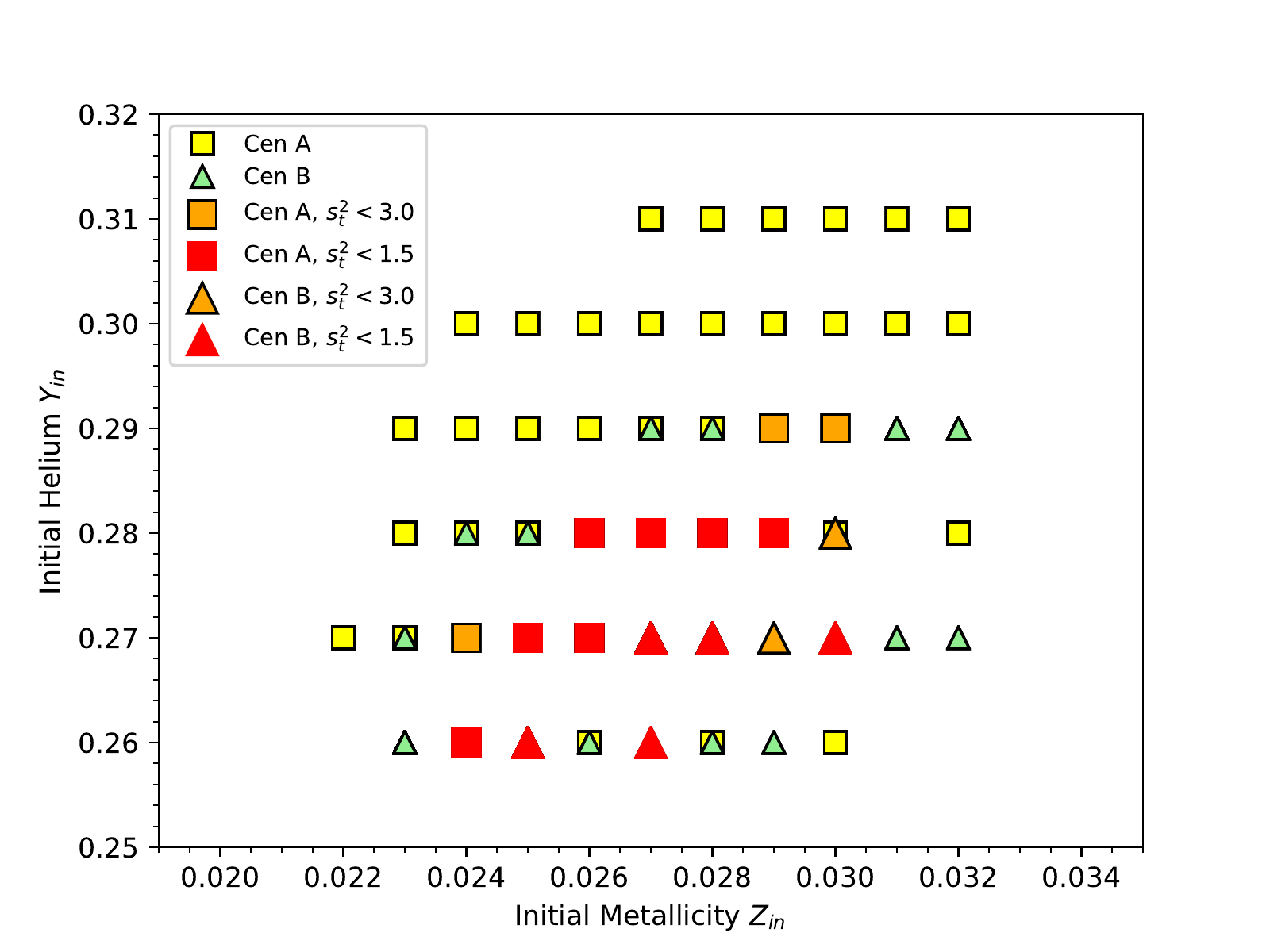} 
\caption{
Helium abundance is shown against metallicity for the whole regime permitted by the classical optimizations, with the seismically allowed regions for {$\alpha$ Cen A and $\alpha$ Cen B} highlighted. {$\alpha$ Cen A} uses square markers and {$\alpha$ Cen B} uses triangles. Preference towards lower $Z_{\text{in}}$ is suggested.
}
\label{heat}
\end{figure}
\begin{figure}
\centering
\includegraphics[width=\columnwidth]{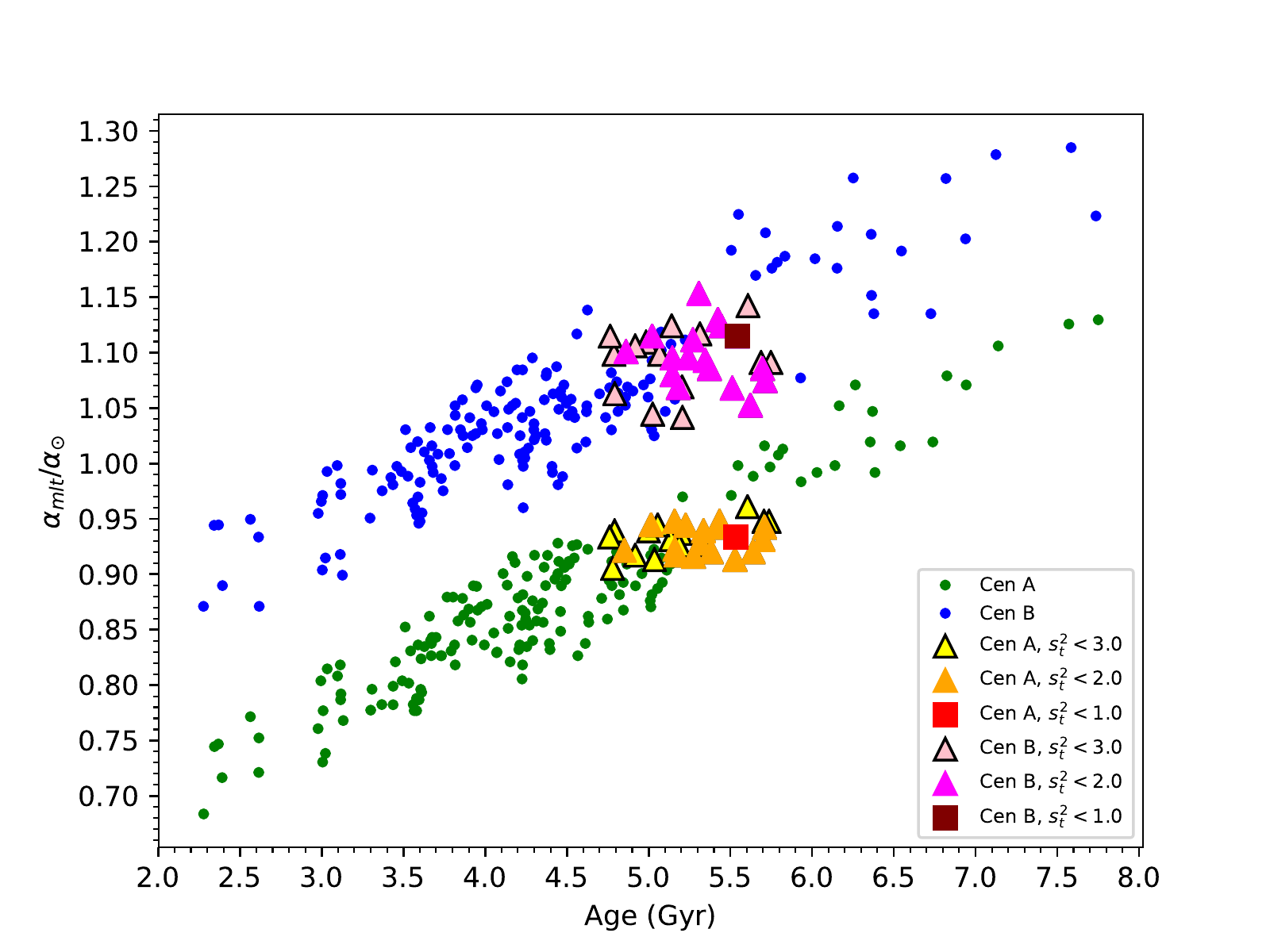} 
\caption{Optimized mixing length is shown as a function of age across the classically permitted regime, with globally permitted parameters highlighted. The inclusion of seismic constraints in a total fit statistic considerably restricts the viable age range.
}
\label{mlt_vs_age}
\end{figure}

Figure \ref{heat} shows $Y_{\text{in}}$ against $Z_{\text{in}}$ and indicates a clear preference for lower helium abundances. Models with $Y_{\text{in}} \ge 0.29$ for {$\alpha$ Cen B} are excluded, and models with $Y_{\text{in}} \le 0.27$ for both stars are preferred. The optimal metallicity is found to fall between $0.024 \le Z_{\text{in}} \le 0.030$, and {$\alpha$ Cen B} prefers a slightly more metal-rich, helium-poor regime.

The galactic chemical enrichment  $\Delta Y / \Delta Z$ can be computed from these data by adopting a value of the primordial helium abundance $Y_p$. We choose $Y_p=0.248$, as this is given in both \citet{Izotov} and \citet{Carigi}. Averaging over all $s^2_{\text{tot,w}}<3$, we obtain $\Delta Y / \Delta Z=0.90$. When separating by star, Cen~A gives $\Delta Y / \Delta Z=1.08$, on average, and {$\alpha$ Cen B} gives $\Delta Y / \Delta Z=0.72$.

These values are low compared to many literature estimates derived from observations, which range anywhere from ${\sim}1.5$ to $\ge5$ depending on the method of inference and region (see e.g.\ \citet{Carigi}, \citet{Gennaro}, and many references therein). \citet{Gennaro}'s Table 3 provides theoretical estimates of $\Delta Y / \Delta Z$ based on standard solar models from different authors and an adoption of $Y_p=0.248$. Our values are {consistent} with these, which are in the range $0.38 \le\Delta Y / \Delta Z\le 0.92$. In addition, \citet{Gennaro}'s Figure 5 shows a distribution of best-fit $\Delta Y / \Delta Z$ occurrences for simulated stars which shifts towards lower values when diffusion is used in stellar models, as is the case in all of ours (to varying degree). The $\Delta Y / \Delta Z$ values suggested by our best fits are consistent with \citet{Gennaro}'s results within $2\sigma$.

Figure \ref{mlt_vs_age} plots age as a function of mixing length for all classically viable models, with the permitted ages corresponding to $s^2_{\text{tot, w}} \le 3$, $s^2_{\text{tot, w}} \le 2$, and $s^2_{\text{tot, w}} \le 1$ highlighted.  While the age is found to be anywhere from 2 to 8 Gyr among classically optimized models, the inclusion of seismic constraints restricts this regime heavily.  When all constraints are considered, allowed ages for the system range between 4.8 and 5.7 Gyr.	Our best estimate for the age of the $\alpha$ Cen system is $5.26 \pm 0.95 $ Gyr, where the uncertainty is the RMS error of the total theoretical range, corresponding to models which fit within ${\sim}3\sigma$. Our $1\sigma$ age estimate is $5.3 \pm 0.3$ Gyr. This is in good agreement with both classical fitting and asteroseismically derived ages in the literature, with \citet{Kim1999}, \citet{Yildiz}, \citet{Bazot2016}, and \citet{Nsamba} all reporting ages ${\sim}4.3$--$5.9$ Gyr.

In summary, Table \ref{averages} gives the average values for mixing lengths, age, and composition for $\alpha$ Cen A and B.  

\begin{table*} 
\centering 
\caption{Average Optimal Parameters by Configuration}
\begin{tabular}{ll  ccc  ccc c}
\\ \hline\hline
Star &
Config &
 &  
Average $\alpha /\alpha_{\odot}$ &
 &
Age &
$Z_{\text{in}}$ &
$Y_{\text{in}}$ &
$\Delta Y / \Delta Z$
\\ \hline
&
&
$s^2_{\text{cl,w}} \le 3.0$ &
$s^2_{\text{t,w}} \le 3.0$ &
$s^2_{\text{t,w}} \le 1.5$ &
&
&
\\ \hline \hline
{$\alpha$ Cen A} &  Standard & 0.899  &  0.940  &	0.934  &  5.157  &  0.028  &
0.280  &  1.140 \\
{$\alpha$ Cen A} &  KS & 0.814  &	0.913  &  0.913  &  5.282  &  0.026  &	0.270  &
0.821 \\
{$\alpha$ Cen A} &  Low Diffusion & 0.886	&  0.937  &  0.933  &  5.452  &  0.028	&
0.278  &  1.094 \\
{$\alpha$ Cen A} &  High Diffusion & 0.881  &  0.925  &  0.944  &	5.119  &  0.026
&
0.275  &  1.033 \\
{$\alpha$ Cen A} &  Overshoot & 0.897  &  0.917  &  0.921	&  5.057  &  0.025  &
0.275  &  1.074 \\
\hline
{$\alpha$ Cen A} & All & 0.884  &$ 0.932  \pm  0.167 $ &  0.931  &  $ 5.261  \pm
0.946 $  & $ 0.027  \pm  0.005 $  & $ 0.277  \pm  0.050 $  & $ 1.079
\pm  0.199 $\\
\hline \hline
{$\alpha$ Cen B} &  Standard & 1.072  &  1.115  &	1.120  &  5.159  &  0.028  &
0.269  &  0.740 \\
{$\alpha$ Cen B} &  KS & 0.962  &	1.056  &  1.068  &  5.268  &  0.026  &	0.265  &
0.647 \\
{$\alpha$ Cen B} &  Low Diffusion & 1.044	&  1.087  &  1.098  &  5.451  &  0.028	&
0.269  &  0.754 \\
{$\alpha$ Cen B} &  High Diffusion & 1.061  &  1.091  &  1.095  &	5.122  &  0.026
&
0.265  &  0.640 \\
{$\alpha$ Cen B} &  Overshoot & 1.077  &  1.098  &  1.101	&  5.060  &  0.026  &
0.265  &  0.647 \\
\hline
{$\alpha$ Cen B} & All & 1.052  & $ 1.095	\pm  0.197 $  &  1.102	&  $ 5.261
\pm  0.946 $ &	$ 0.027  \pm  0.005 $ &  $ 0.268  \pm  0.048 $ &  $
0.715 \pm  0.133 $\\
\hline
\end{tabular}
\tablecomments{Ages, helium abundances, metallicities, and enrichment
fractions are averaged over models with $s^2_{\text{t,w}}\le
3$. $Y_p=0.248$ is adopted for enrichment calculations.
}
\label{averages}
\end{table*}

\section{Summary and Conclusions}
\label{conc}
We have computed optimal parameters for grids of stellar models tailored to reproduce the observed features of $\alpha$ Cen A and B as provided by \citet{Kervella17}, \citet{PortoDeMello08}, \citet{Thoul03}, \citet{deMeulenaer}, and \citet{Kjeldsen}. We consider five different sets of assumptions about the modeling physics, including two analytical atmospheric boundary prescriptions, three diffusion prescriptions, and the activation (or not) of convective core overshoot, where $\alpha_{\text{ovs}}=0.1$.

For all configurations, we find input parameter combinations which produce good agreement with classical and binary observational constraints. Overall, models with standard physics (normal diffusion, Eddington model atmosphere and no convective core overshoot) produce the best fits, followed by models which vary the efficiency of diffusion, and models which allow for some amount of convective core overshoot. Models which use the \citet{KrishnaSwamy} scaled solar model atmosphere yield the worst fits to the data.

We report best-fitting parameters for the $\alpha$ Cen system:	
 $\alpha_{\text{MLT,A}} / \alpha_{\odot} = 0.932 \pm 0.17$; 
 $\alpha_{\text{MLT,B}} / \alpha_{\odot} = 1.095 \pm 0.20$;
$t=5.26 \pm 0.95$ Gyr;
$\bar{Z_{\text{in}}} = 0.027 \pm 0.005$;
$\bar{Y_{\text{in}}} = 0.273 \pm 0.035 $; 
and $\Delta Y / \Delta Z = 0.90 \pm 0.12$. The theoretical uncertainties on these quantities are $3\sigma$, given by the RMS errors computed from 31 models found to agree with all classical, binary, and observational constraints within $s^2_{\text{tot,w}}\le3$.

The globally optimized mixing lengths of $\alpha$ Cen A and~B  are found to converge to roughly the same non-solar values regardless of choices in input physics. The values are well constrained and consistent with evidence that $\alpha_{\text{MLT}}$ is a function of several physical attributes, rather than a static value \citep{Liu2018, Joyce2018, Viani,Tayar,Creevey17, Bonaca}. The finding $ \alpha_{\text{MLT,A}} < \alpha_{\odot} < \alpha_{\text{MLT,B}}$ in particular is consistent with \citet{Joyce2018} in reproducing the need for higher mixing lengths for stars with higher mass.

Seismic parameters are found to restrict the range of allowed mixing lengths over classical values and to raise them on average. They are also found to constrain model ages considerably. The fitted ages found with this method are in excellent agreement with literature ages \citep{Kim1999,Yildiz,Bazot2016, Nsamba}. Chemical enrichment estimates based on our models are low, but in good agreement with other theoretical values quoted for stellar models which consider heavy element diffusion and precise chemical abundance prescriptions \citep{Gennaro, Asplund09}.

Of 31 viable models, two find core convection in Cen~A, and both use enhanced diffusion and $\alpha_{\text{ovs}}=0.1$ (for Cen~A) in their modeling physics. Rather than make statistical estimates on the likelihood of a convective core in $\alpha$ Cen A (as in \citealt{Bazot2016} and \citealt{Nsamba}), our analysis focuses on considering the widest range of physical possibilities and satisfying as many constraints as possible. Our work suggests that if {$\alpha$ Cen A} is found to have a convective core, modifications to standard physical prescriptions (such as enhancing diffusion) would be necessary to model {$\alpha$ Cen A} appropriately.

With the recent availability of high-quality classical and asteroseismic observations, it is becoming well-understood that use of the solar mixing length in non-solar models no longer constitutes good stellar modeling.
{In the near future, NASA's Transiting Exoplanet Survey Satellite (\textit{TESS}) mission \citep{TESS} will provide asteroseismic measurements for hundreds of thousands of stars--- $\alpha$ Centauri among them. 
Improved seismic observations of $\alpha$ Cen A and B could lead to significantly tighter constraints on the stellar models, but the separation of $\alpha$ Cen A and B will be less than 6'' for the next few years \citep{Kervella16}, making them challenging targets for missions with low angular resolution. If blending can be resolved, \textit{TESS} could potentially provide enough new frequency measurements to incorporate additional seismic ratios (i.e.\ besides $r_{02}$; see \citealt{RoxVoronstov}) in our optimization criteria, yielding higher precision estimates for the mixing lengths in $\alpha$ Cen A and B and the age of the system.}

By calibrating the mixing length parameter under the requirement that our models of $\alpha$ Cen satisfy all known observed features---including those for $\alpha$ Cen B---{and by incorporating the best empirical data available, we move closer} toward solar-precision modeling of other stars.

\acknowledgements
This work is supported by grant AST-1211384 from the National Science Foundation. M.\ Joyce would like to thank L\'aszlo M\'oln\'ar for helpful discussions about asteroseismology and John Bourke for typesetting. We would like to thank Konkoly Observatory, the South African Astronomical Observatory, and the University of Cape Town for resource support during this study.

\bibliographystyle{apj}
\bibliography{alphacen_mltV8}

\end{document}